\documentclass[pre,longbibliography,showpacs,twocolumn,floatfix,superscriptaddress]{revtex4-1}

\usepackage{graphicx}
\usepackage{epstopdf}
\usepackage{dcolumn}
\usepackage{array}
\usepackage{amsmath}
\usepackage{amssymb}
\usepackage{units}
\usepackage{xcolor}
\usepackage{bm}
\usepackage{setspace}
\usepackage{color}
\usepackage{textcomp}

\definecolor{dark_green}{RGB}{0, 120, 0}

\draft

\begin{document}
\title{Electrified Cone Formation in Perfectly Conducting Viscous Liquids:\\ Self-Similar Growth Irrespective of Reynolds Number}
\author{Theodore G. Albertson and Sandra M. Troian}
\homepage[Corresponding author (URL/email) :~~]{www.troian.caltech.edu; stroian@caltech.edu}
\affiliation{California Institute of Technology, T. J. Watson Sr. Laboratories of Applied Physics, 1200 E. California Blvd., MC 128-95, Pasadena, CA 91125, USA}

\date{August 8, 2019}

\begin{abstract}
Above a critical field strength, the free surface of an electrified, perfectly conducting viscous liquid, such as a liquid metal, is known to develop an accelerating protrusion resembling a cusp with a conic tip. Field self-enhancement from tip sharpening is reported to generate divergent power law growth in finite time of the forces acting in that region. Previous studies have established that tip sharpening proceeds via a self-similar process in two distinct limits - the Stokes regime at $\textsf{Re}=0$ and the inviscid regime $\textsf{Re} \to \infty$. Using finite element simulations to track the acceleration of an electrified protrusion in a perfectly conducting Newtonian liquid in vacuum held at constant capillary number, we demonstrate that the conic tip \textit{always} undergoes self-similar growth irrespective of Reynolds number. The computed blow up exponents at the tip for the terms in the Navier-Stokes equation and interface normal stress condition reveal the different forces at play as $\textsf{Re}$ increases. Rescaling of the tip shape by the capillary stress exponent yields excellent collapse onto a universal conic tip shape with interior half-angle dependent on the magnitude of the Maxwell stress. The rapid acceleration of the liquid interface also generates a thin surface boundary layer with very high local strain rate. Additional details of the modeled flow, applicable to cone growth in systems such as liquid metal ion sources, help dispel prevailing misconceptions that dynamic cones resemble conventional Taylor cones or that viscous stresses at finite $\textsf{Re}$ can be neglected.\\

Credit line: This article has been submitted to the \textit{Physics of Fluids} - a link to the final version will be provided once published.
\end{abstract}

\maketitle

\section{Introduction}
In a short and influential paper in 1884, Lord Rayleigh demonstrated by a simple argument based on linear stability analysis why a charged spherical mass of liquid acted upon solely by electrical and capillary forces represents a state of unstable equilibrium. Maxwell's equations had only been developed a couple decades earlier so Rayleigh's hydrodynamic analysis was especially novel and insightful. He predicted too that the instability would cause the spherical mass to distort into a peaked structure that would emit a fine jet due to the growing imbalance between the destabilizing electrical pressure and the stabilizing capillary pressure. Several years later in 1890, J. Larmor became interested in the electrification of capillary waves on a liquid surface \cite{Larmor:pcps1890}. Seemingly unaware of Rayleigh's work, he examined the linear stability of an electrified liquid surface by appealing to the unsteady form of Bernoulli's equation from which he derived the dispersion relation correlating the rate of growth to the wavelength of ripples along the interface of a charged liquid film. These two early studies thereby revealed the ease with which an electrified liquid surface can develop and sustain protrusions which narrow and sharpen in time.

\subsection{Instability analyses by Tonks and Frenkel}
For decades, this phenomenon remained mostly a curiosity until 1935 when L. Tonks \cite{Tonks:pr1935}, a researcher at General Electric who was investigating ion emission from a solid metal electrode, turned his attention to sparking and charge emission in liquid metals. Tonks closely examined the liquid distortion process and electrical breakdown of a liquid mercury surface in close proximity to a counter electrode for field strengths approaching several megavolts/cm. In particular, he noted that a planar liquid metal surface rapidly generated an accelerating protrusion whose narrowing shape concentrated the local disruptive electric field so as to generate a liquid cusp capable of ion emission. His studies eventually confirmed that field induced ion emission from liquid metals occurs at lower field strengths than required for solid metals. Informed by these observations, he carried out a number of calculations which resulted in expressions and estimates relating the amplitude of the distortion of an electrified liquid surface to the minimal field strength required to induce a sustained deformation, and the typical time required for interface rupture to occur resulting in ion emission.  Tonks emphasized two key questions that continued to drive his work - namely, what was the relation between the applied field strength, the distortion amplitude and the pressure at the tip of the accelerating protrusion, and how fast did this protrusion advance in response to the pressure at the tip? Realizing that a complete and rigorous solution to this problem was rather complicated, he instead developed an insightful approximate treatment that revealed critical aspects of the distortion runaway process. Shortly after Tonk's work was published, J. (Ya. I.) Frenkel \cite{Frenkel:zetf1935,Frenkel:pzs1936} provided a more general derivation of this dynamic process by investigating the velocity of surface waves triggered by an applied electric field. Such periodic distortion of the surface of a perfectly conducting liquid for sufficiently high electric field strengths is now known as the Larmor-Tonks-Frenkel instability. Despite the number of studies and contributions to this field during the past century, the two questions originally posed by Tonks have not yet been completely answered.

The investigations by Larmor and Frenkel relied on a stability analysis of capillary waves on a perfectly conducting liquid subject  to an electric field, through which they uncovered that the wave frequency becomes complex valued once the critical field strength for sustained distortion is exceeded. By contrast, Tonks focused on the dynamic evolution of a single protrusion emanating from a flat liquid surface with an initial small bump modeled as a spherical cap. His analysis revealed that beyond a critical field strength, the pressure at the cap apex cannot support equilibrium since the electric pressure at the tip - now known as the Maxwell pressure or Maxwell stress - increases as the square of the apex height. This quadratic dependence cannot be balanced by the restoring hydrostatic or capillary pressures, which at best scale linearly or inversely with apex height, respectively. Therefore, any small increase in the apex amplitude simply increases the net local unbalanced pressure. This strong imbalance causes a runaway process by which the distortion progressively increases in amplitude. By relying on various approximations for the initial conditions and simplified functional forms for other quantities, he was successful in deriving an equation for the acceleration of an eccentric liquid shape with time. Remarkably, Tonks also seemed to realize that the growth of the liquid tip proceeded by a self-similar process, as is apparent from his sketch of the time sequence of an evolving interface shape in Fig. 8 of Ref. [\onlinecite{Tonks:pr1935}].

\subsection{Experimental studies by Krohn and co-workers}
Interest in this phenomenon began to spread and by the early 1960's, V. E. Krohn \cite{Krohn:paa1961} and co-workers at TRW recognized that the spontaneous process by which an electrified liquid can undergo fission to produce charged droplets and ions could be of enormous relevance to the burgeoning field of electrostatic propulsion \cite{Perel:jsr1971,Bailey:jsr1972,Zafran:jsr1973,Huberman:jsr1974} and in particular, development of so-called ``space thrusters''. While other researchers focused on colloid thrusters based on electrospray from dielectric liquids, Krohn instead concentrated on cusp formation in liquid alkali metals. Within a few years, he  successfully demonstrated ion emission from small liquid droplets of cesium, gallium and mercury supported on the tip of a tungsten needle maintained at a surface potential of a few kilovolts. By 1975, Krohn was able to demonstrate stable emission lasting several hours, a remarkable technological achievement which essentially paved the way for a number of technologies based on liquid metal ion sources (LMIS). By the late 1970's, researchers \cite{Krohn:rsi1972,Seliger:apl1979} recognized the potential of LMIS for high-resolution ion microprobe analysis, micro-milling and micro-implantation. Attention therefore soon turned to identification  and control of those operational parameters affecting the diameter of the ion source area, the maximum brightness achievable and the energy spread of the ion beam. Significant industrial support led to rapid and far reaching developments in this field, so much so that LMIS now forms the basis for instrumentation ranging from high resolution focused ion beam systems for enhanced etching and deposition, to scanning ion microscopy, micro-milling, ion mass spectrometry, implantation and lithography \cite{Orloff:2003}. Equally exciting are numerous current developments aimed at space micropropulsion systems based on LMIS such as field emission electric propulsion \cite{Wright:pas2015,Bharti:irjet2017} (and references therein).

\subsection{NASA interest in liquid metal ion sources for space micropropulsion}
The most recent 2015 NASA Technology Roadmap \cite{NASA15} identifies in-space micropropulsion technologies based on micro-electrospray propulsion as key enabling technologies, defined as small scale systems which can deliver thrust using a conductive fluid and electrostatic fields to extract and accelerate ions, charged droplets, molecules or clusters of molecules depending on whether the propellant is a liquid metal or ionic fluid. These systems are intended to provide high specific thrust levels characterized by low power, small form factors and extended lifetimes. The ultimate goal is to extend current know-how gained from the fabrication and manufacture of microelectromechanical systems (MEMS) to micropropulsion platforms consisting of hundreds to thousands of needle microarrrays externally wetted by a film of liquid metal electrified on demand to emit charged particles. NASA envisions that such miniaturized systems will deliver thrust levels suitable for precision attitude, rendezvous and docking maneuvers without the use of reaction wheels. Such capability is anticipated to advance formation flight and attitude actuation for small distributed spacecraft \cite{You:2018}.

The fluid dynamical process leading to cusp formation in free surface, perfectly conducting, viscous liquids has remained a key problem in the field of electrohydrodynamics for decades. Prior studies of cusp formation have focused almost exclusively on two asymptotic limits - the Stokes flow regime at $\textsf{Re}=0$ and the inviscid regime at $\textsf{Re} = \infty$. Since the dynamic behavior of rapidly accelerating, electrified viscous fluids is a rather complex process, only these two asymptotic limits have so far yielded analytic insight by revealing which forces at the liquid tip are responsible for self-similar growth. Natural questions to ask at this point are whether the self-similar evolution of the electrified tip persists for all Reynolds numbers, how does the tip shape depend on the Reynolds number $\textsf{Re}$ and capillary number $\textsf{Ca}$, and what are the exponents characterizing the divergent behavior at the tip for the different forces involved just prior to ion emission.

Using finite element simulations to track the acceleration of an initially tiny electrified protrusion in a perfectly conducting Newtonian liquid in vacuum held at constant capillary number, we demonstrate that the conic tip \textit{always} undergoes self-similar growth irrespective of Reynolds number. These axisymmetric simulations allow high resolution inspection of the acceleration and sharpening of an initial rounded tiny protrusion on an otherwise flat liquid layer. The computed blow up exponents at the tip for the terms in the Navier-Stokes equation and interface normal stress condition reveal the different forces at play as $\textsf{Re}$ increases. Rescaling of the tip shape by the capillary stress exponent yields excellent collapse onto a universal conic tip shape with interior half-angle dependent on the Maxwell stress magnitude. Additional details pertaining to the importance of viscous stresses even at higher Reynolds number as well as the formation of a surface boundary layer characterized by a very high local strain rate are also presented.

\section{Notable Studies of Electric Field Induced Cuspidal Formation in Liquids}
\label{Background}

\subsection{Electro-stationary state known as a Taylor cone}

The analyses of Rayleigh \cite{Rayleigh:pm1882} and Larmor \cite{Larmor:pcps1890} made evident that above a critical surface charge or above a critical field strength, a perfectly conducting mass of liquid must incur deformation represented by a state of unsteady equilibrium. In 1964, Taylor \cite{Taylor:prsla1964} examined what would be the equilibrium shape which would everywhere on the surface allow for a perfect balance between the local capillary pressure and local Maxwell pressure. This static shape was found to be a cone of constant slope with an interior half-angle of $49.3^\textrm{o}$, now known as the Taylor angle. While the two opposing normal stresses are everywhere in mechanical balance, however, they undergo divergence at the conic apex, a feature which troubled subsequent researchers. Over the years, this result has tended to cause considerable confusion between Taylor's stationary configuration and the more prevalent dynamic conic shapes, such as conic cusps, which represent non-equilibrium configurations arising from the increasing imbalance in forces at the liquid tip. Some authors \cite{Orloff:2003} have attempted to clarify this issue by referring to dynamic protrusions as ``Gilbert-Gray cone-jets''. Gilbert \cite{Gilbert:petrus1600} first reported the attraction of water to a charged piece of amber with subsequent formation of a conic protuberance from whose vertex was emitted a very fine jet and Gray \cite{Gray:prsl1731} reported similar jet formation not only in water but liquid mercury as well.

\subsection{Zubarev analysis of self-similar conic cusp formation in the inviscid limit ($\textsf{Re}=\infty$)}
In a seminal paper in 2001, Zubarev \cite{Zubarev:jetp2001} analyzed the dynamic distortion accompanying the response of a perfectly conducting layer of liquid to an initially uniform electric field. His asymptotic analysis of the unsteady Bernoulli equation applied to the moving liquid interface was conducted under the assumptions of incompressible, inviscid and irrotational flow. Key to this analysis was the assumption that the electric field strength near the advancing tip rapidly and appreciably exceeds the applied external field value such that local conditions prevail and the governing electric field strength decreases with distance from the apex. The external field uniformity condition could also then be replaced with a far field boundary condition specifying vanishing electric field strength at infinity. This approach allowed him to analyze the fluid motion near the singular apex without reference to any particular geometry, ultimately revealing the existence of self-similar solutions for the velocity potential, electric potential and fluid interface shape when rescaled in time by the variable $\tau = T_\textsf{C} - T >0$ where $T_\textsf{C}$ is the dimensionless collapse time and $T$ the dimensionless time. We next outline Zubarev's analysis leading to this prediction of self-similar growth and the exponents characterizing the divergence at the conic tip of the Maxwell pressure, capillary pressure and kinetic energy density.

We note that for the remainder of this work unless otherwise designated, dimensional quantities are designated by lower case letters and dimensionless quantities by upper case letters. An exception to this is the electric field vector represented in dimensional form by $\bm{\hat{E}}$ and in dimensionless form by $\bm{E}$. The vector $\bm{\hat{n}}$ denotes a unit normal vector at the moving interface pointing outward from the liquid domain. Bold faced variables are used to denote vector and tensor quantities.

Zubarev examined the evolution and growth of an electrified axisymmetric protrusion in a perfectly conducting inviscid fluid. Here, the interface function $z=h(r,z,t)$ denotes the boundary separating the liquid and vacuum regions. The vacuum domain supports an electric field distribution $\bm{\hat{E}}(r,z,t)=-\nabla \phi (r,z,t)$, where the electric potential $\phi(r,z,t)$ satisfies the Laplacian condition $\nabla^2\phi(r,z,t)=0$. The velocity field within the liquid region is described by $\bm{u}(r,z,t)=\nabla \psi(r,z,t)$ where the velocity potential $\psi(r,z,t)$ satisfies $\nabla^2\psi(r,z,t)=0$ for an incompressible liquid subject to irrotational flow. Given that the liquid is also perfectly conducting, its interior domain defines a Gaussian volume devoid of electric field and therefore only the moving interface experiences an electrical normal force which pulls liquid toward the vacuum domain. Conservation of mass and momentum are enforced through the unsteady form of Bernoulli's equation for inviscid irrotational flow given by
\begin{align}
\rho & \left(\psi_t+\frac{|\nabla\psi|^2}{2}\right) = \nonumber \\
& \underbrace{\frac{\epsilon_0|\nabla\phi \cdot \bm{\hat{n}}|^2}{2}}_\text{Maxwell pressure} + \underbrace{\frac{\gamma}{(1+h_r^2)^{1/2}}\left(\frac{h_{rr}}{1+h_r^2}+
\frac{h_r}{r}\right)}_\text{Capillary pressure}
\label{eqn:ZubarevBernoulli}
\end{align}
for a liquid with constant density $\rho$ and constant surface tension $\gamma$. The underbrace terms represent the opposing local Maxwell and capillary pressures acting in the direction normal to the moving interface. Zubarev non-dimensionalized the governing variables and equations by introducing the following scalings:
\begin{eqnarray}
\label{eqn:StartNondim}
(R,Z,H) &=& (r,z,h) \times \frac{\epsilon_o}{\gamma} \hat{E}^2_n\\
T &=& t \times \frac{\epsilon_o^{3/2}}{\gamma\rho^{1/2}} \hat{E}^3_n\\
\Phi &=& \phi \times \frac{\epsilon_o}{\gamma} \hat{E}_n\\
\Psi &=& \psi \times \left(\frac{\rho \epsilon_o}{\gamma^2}\right)^{1/2} \hat{E}_n ~,
\label{eqn:EndNondim}
\end{eqnarray}
where $\hat{E}_n = |\bm{\hat{E} \cdot \hat{n}}|$. The boundary conditions used to solve Eq. (\ref{eqn:ZubarevBernoulli}) and the corresponding Laplacian fields were chosen to be
\begin{eqnarray}
\label{eqn:StartBCs}
\Phi [Z=H(R,T)] &=& 0\\
\lim_{\textbf{R}\to\infty} (\Phi^2_R + \Phi^2_Z) &=& 0 ~~~~~~~~~~Z > H(R,T)\\
\lim_{\textbf{R}\to\infty} (\Psi^2_R + \Psi^2_Z) &=& 0 ~~~~~~~~~~Z < H(R,T)\\
\Phi_R (R=0,Z,T)&=& 0 \\
\Psi_R (R=0,Z,T)&=& 0 \\
H_R(R=0,Z,T) &=& 0\\
\label{eqn:EndBCs}
H_T - \Psi_Z + \Psi_R H_R&=&0
\label{eqn:ZubarevKinematic}
\end{eqnarray}
where $\textbf{R}=R \bm{\hat{e}}_R + Z \bm{\hat{e}}_Z$. A scaling analysis of the governing Bernoulli equation and boundary conditions revealed the existence of self-similar solutions, here indicated by a tilde, according to which
\begin{align}
\label{eqn:StartSelfSim}
\left[\widetilde R,\widetilde Z,\widetilde H(\widetilde R)\right] &= \frac{\left[R,H(R,T)\right]}{\tau^{2/3}}\\
\widetilde \Phi(\widetilde R,\widetilde Z)&=\frac{\Phi(R,Z,T)}{\tau^{1/3}} ~~~~~~~~~~ \widetilde Z > \widetilde H(\widetilde R)\\
\widetilde \Psi(\widetilde R,\widetilde Z)&=\frac{\Psi(R,Z,T)}{\tau^{1/3}} ~~~~~~~~~~ \widetilde Z < \widetilde H(\widetilde R)\\
\textrm{where}\,\,\,\tau&=T_\textsf{C}-T \,
\label{eqn:EndSelfSim}
\end{align}
where $T_\textsf{C}$ denotes the dimensionless asymptotic collapse time when the interface shape $\widetilde H$ assumes a universal form.

Zubarev constructed self-consistent solutions to these equations by matching the self-similar solutions in the inner region at the liquid apex to the far field solutions describing a conventional stationary Taylor cone with interior half-angle $\theta_o = 49.3^\textrm{o}$. To leading order as $\bar{R}=(\widetilde R^2+\widetilde Z^2)^{1/2}\to\infty$ (or equivalently $\tau \to 0$ in self-similar coordinates), he found that the asymptotic functions scaled as
\begin{align}
\label{eqn:ZubHilite}
\widetilde \Psi(\widetilde R,\widetilde Z)&\propto s/\bar{R}\\
\widetilde \Phi(\widetilde R, \widetilde Z)&\propto \frac{1}{c_o} \left[2 \bar{R} (s_o-s)\right]^{1/2} P_{1/2}(\cos\theta)\\
\widetilde H(\widetilde R)&=-s_o \widetilde R\\
\end{align}
where
\begin{align}
P_{1/2}(\cos\theta_o)&=0\\
s_o &= - \cot \theta_o \\
c_o &= \left(\frac{d P_{1/2}(\cos\theta)}{d \theta}\right)_{\theta = \theta_o}~,
\end{align}
where $\theta=\arctan(\widetilde R/\widetilde Z)$ denotes the polar angle in spherical coordinates, $P_{1/2}(\cos\theta)$ denotes the Legendre polynomial of order 1/2, and $s$ is a constant such that $ 0 < s < s_o$. As evident from Eq. (\ref{eqn:ZubHilite}), the solution corresponds to fluid motion that is spherically symmetric such that liquid is transported to the conic apex by streamlines which are always tangent to the moving interface. The point $(\widetilde R,\widetilde Z)=(0,0)$ therefore acts as a point sink, a feature Zubarev highlighted.

These self-similar solutions describe divergent behavior of the  Maxwell pressure $P_M$, capillary pressure $P_C$ and flow speed $|\bm{U}|$ at the advancing conic apex according to which
\begin{align}
P_M \sim P_C &\sim \tau^{-2/3} \\
|\bm{U}| &\sim \tau^{-1/3} ~.
\label{eqn:ZubPVScalings}
\end{align}
Through this analysis, Zubarev was the first to report that a  perfectly conducting liquid subject to inviscid, irrotational and incompressible flow set by local conditions due to Maxwell and capillary forces undergoes local self-similar evolution that in the far field smoothly matches onto the conventional stationary Taylor cone shape. Zubarev described the resulting interface shape as a conic cusp.

\subsection{Self-Similar Cusp Formation in the Viscous Regime}
\subsubsection{Numerical Studies by Suvorov, Litvinov and Zubarev}
Around the time Zubarev reported that the inviscid limit supports self-similar growth of the conic tip, Suvorov and Litvinov \cite{Suvorov:jpd2000} conducted a computational study based on the axisymmetric geometry shown in Fig. \ref{fig:SimGeometry} to better understand the influence of viscous effects. They employed a cell and marker technique with a coordinate transformation method for adaptive grid generation \cite{Belozerkovskii:1994}, which allowed smooth and stable evolution of the fluid interface despite development of significant and rapidly increasing curvature at the conic apex. The numerical technique they implemented prevented onset of spurious surface oscillations near the region of high curvature, a problem which had plagued previous researchers \cite{Cui:jvstb1988}.

According to Fig. \ref{fig:SimGeometry}, the axisymmetric volume of interest was partitioned into two contacting domains - an upper domain comprising the vacuum region subject to an initial uniform electric field of strength $E_o= 2.4 \times 10^ 8$ V/m and a lower domain comprising a perfectly conducting, incompressible Newtonian liquid. Given the large disparity in density and viscosity, the hydrodynamics of the vacuum domain was neglected. The normal component of the electric field strength along the vacuum-liquid interface (held at constant electric potential) was updated in time according to its spatiotemporal location. The Navier-Stokes equation for the liquid domain was solved self-consistently subject to the kinematic boundary condition governing the motion of the moving interface. The initial thicknesses of the vacuum and liquid domains, along with their radial extent, were chosen to be 20 $\mu$m. The material constants of the liquid were chosen match those of liquid mercury. The initial condition for the interface represented a flat film containing a tiny Gaussian bump centered at the origin of the form $h(r,t=0) = a \exp[-(r^2/\lambda^2)]$ where $a= 0.4 \, \textrm{or} \, 4.0 \, \mu\textrm{m}$ and $\lambda = 4 \, \mu\textrm{m}$. The chosen dimensions and material constants therefore corresponded to laminar flow conditions at the value $\textsf{Re} = 717$.

For an initial Gaussian bump of fixed lateral extent, the simulations showed that a thicker liquid layer tended to produce a more rounded protrusion advancing rapidly in time, while a thinner liquid layer produced a more pronounced cuspidal shape that advanced more slowly, presumably due to stronger viscous effects. Suvorov and Litvinov called these structures \textit{dynamic} Taylor cones to emphasize that such shapes do not resemble conventional stationary Taylor cones. Instead, the initial protrusion was observed to deform into a cuspidal formation with a cone-like tip terminated by a rapidly shrinking spherical cap. Suvorov and Litvinov also carried out simulations in which the viscous term in the normal stress boundary condition was altogether omitted. They concluded from these additional studies that the viscous normal force at the advancing tip is always positive - that is, it too acts in the same direction as the capillary force but is much smaller in magnitude. They reported that the resulting shapes obtained with or without the viscous normal stress were very similar but did not include any comparative figures.

The numerical scheme upon which this worked was based, which allowed stable interface growth over half a decade in time, successfully  capture divergent behavior at the conic apex of such quantities as the curvature radius, electric field strength and axial velocity. These divergent quantities led in turn to runaway behavior in the  Maxwell pressure, capillary pressure, viscous normal stress and kinetic energy density. Their study, conducted at a fixed value  $\textsf{Re} = 717$, demonstrated that the (negative) Maxwell normal stress pulling upward on the liquid apex rapidly overtakes the (positive) normal stresses due to capillary and viscous forces. Suvorov and Litvanov concluded by urging that more such computational studies be conducted to assist theorists in developing analytical treatments for the dynamic behavior of electrified liquids used in liquid metal ion sources. While their work established key features characterizing the runaway cusp formation process in viscous fluids, their results were restricted to a single value of $\textsf{Re}$. They also did not report the actual values obtained for the exponents characterizing the various divergent forces at the accelerating tip.
Soon thereafter, there followed similar computational studies by Suvorov \cite{Suvorov:surf2004} and Suvorov and Zubarev \cite{Suvorov:jpd2004} simulating the behavior of liquid gallium at the melting point at flow conditions corresponding to a smaller value of the Reynolds number. In these studies, the viscous normal stress term in the interface boundary condition was altogether omitted based on earlier indications \cite{Suvorov:jpd2000} that the viscous term in the normal stress boundary condition tended to be much smaller than the capillary term. In these simulations, the initial thicknesses of the vacuum and liquid domains, along with their radial extent, were reduced to 5 $\mu$m while the initial applied electric field strength value was increased to $E_o= 4.8 \times 10^ 8$ V/m. The initial condition for the interface was again chosen to be a flat film with a tiny Gaussian bump centered at the origin of the form $h(r,t=0) = a \exp[-(r^2/\lambda^2)]$ where $a= 0.1 \,\mu\textrm{m}$ and $\lambda = 1 \,\mu\textrm{m}$. These Gaussian parameters, corresponding to an initial radius of curvature of the liquid protrusion equal to $\lambda^2 / 4 a = 2.5 \, \mu \textrm{m}$, were chosen to approximate the radii of supporting needles typically used in liquid metal ion sources. The chosen dimensions and material constants in these studies therefore corresponded to a value  $\textsf{Re} = 187$.

\begin{figure}[!]
\includegraphics[width=8.5cm]{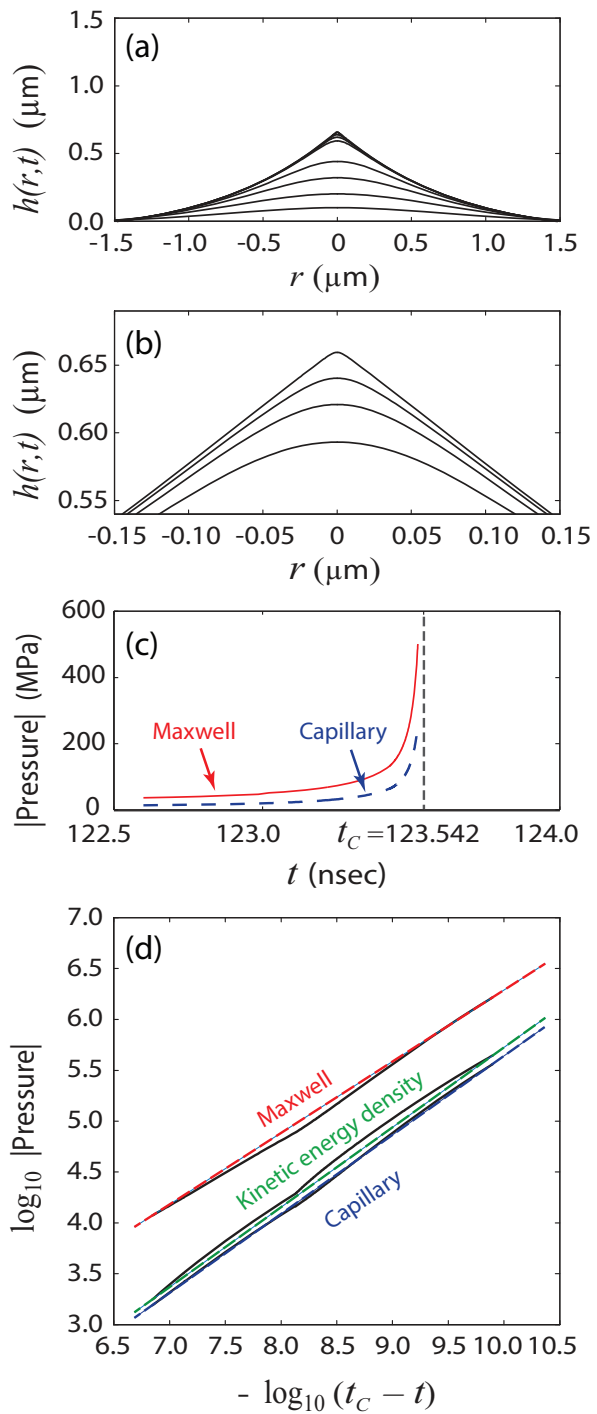}
\caption{Results from Ref. [\onlinecite{Suvorov:surf2004}] showing self-similar formation of an electrified liquid cusp in a thin flat film of gallium (5 ~ $\mu \textrm{m}$ in thickness) with a tiny Gaussian bump of initial amplitude $0.1 ~\mu \textrm{m}$ centered about the origin. Initial electric field strength $E_o = 4.8 \times 10^8 \textrm{V/m}$. (a) Interface shapes shown at times $t=$ 0, 81.2, 104.5, 116.1, 122.6, 123.1, 123.4 and 123.5 nsec. (b) Magnified view of conic tip at $t=$ 122.6, 123.1, 123.4 and 123.5 nsec. (c) Divergent behavior of the (positive) Maxwell pressure and (negative) capillary pressure at the conic apex upon approach to the collapse time $t_\textsf{C}=123.542$ nsec. (d) Log-log plot showing power law growth of the (dimensionless) Maxwell pressure, kinetic energy density and capillary pressure on approach to the collapse time.}
\label{fig:SuvorovPlots}
\end{figure}

The numerical results by Suvorov \cite{Suvorov:surf2004} are shown in Fig. \ref{fig:SuvorovPlots}. Only this set of figures will be  discussed here since the results presented in Ref. [\onlinecite{Suvorov:jpd2004}] are similar. Snapshots of the interface shape depicted in Figs. \ref{fig:SuvorovPlots} (a) and (b) illustrate both the far and near field views of an evolving protrusion. These images reveal how the rapidly accelerating interface undergoes self-similar collapse to an apical shape resembling a sharpening conic tip with a rapidly decreasing radius of curvature. The curves in Fig. \ref{fig:SuvorovPlots}(c) illustrate  blow up behavior at the apical tip in the magnitudes of the Maxwell and capillary pressure as the system approaches the collapse time of 123.542 nsec. Fig. \ref{fig:SuvorovPlots}(d) represents a log-log plot of the magnitude of the apical values of the Maxwell pressure, capillary pressure and kinetic energy density upon approach to the collapse time. The solid lines shown were reported to yield exponent values $\beta$ corresponding to divergent behavior $(t_\textsf{C} - t)^{-\beta}$ of the Maxwell pressure, capillary pressure and kinetic energy density of 0.67, 0.68 and 0.63, respectively. The dashed lines we have superposed represent the corresponding exponent values 0.70, 0.77 and 0.78, which provide a closer fit to Suvorov's data.

\subsubsection{More recent numerical studies of conic formation in electrified liquids}

More recently, Collins et al. \cite{Collins:natp2008} used the finite element method with elliptic mesh generation and adaptive time integration to simulate liquid cone formation in perfectly conducting, perfectly insulating and leaky dielectric liquids. A goal of their study was to understand why electrohydynamic tip streaming resulting in the emission of a thin fluid jet from the conical tip does \textit{not} occur in perfectly conducting or perfectly insulating liquids. Their simulations of perfectly conducting liquids also revealed tip sharpening with corresponding divergent behavior in the Maxwell, capillary and viscous normal stress at the apex, in agreement with previous studies \cite{Suvorov:jpd2000,Suvorov:jpd2004,Suvorov:surf2004} - however, they did not extract the corresponding values of the blow up exponents. Most interestingly, this study confirmed that tip-streaming requires the existence of interfacial tangential stresses near the conic tip, which are ultimately responsible for the cone-jet transition observed. These numerical simulations successfully captured for the first time the various scaling laws governing the size of droplets formed from steady cone-jets as first predicted by de la Mora and Loscertales and Ga\~{n}\'{a}n-Calvo \cite{delaMora:jfm1994,GananCalvo:prl1997}.

Burton and Taborek \cite{Burton:prl2011} have since conducted detailed boundary integral simulations to examine Coulombic fission of charged inviscid droplets. Their studies have revealed how small disturbances cause rapid elongation of an initial slightly prolate ellipsoidal droplet to produce polar cap regions resembling conic tips, which they have coined lemon-shaped drops. For perfectly conducting liquids, their results show how conic tips undergo self-similar sharpening due to local field self-enhancement with a corresponding blow up exponent of 2/3 for the capillary pressure and 1/3 for the charge density. Remarkably, they were able to confirm self-similarity over twelve decades in time - an extraordinarily long period - requiring stable integration able to resolve high resolution behavior at the advancing tip. Incidentally, they discovered that during conic sharpening, the interior half-angles never equal the Taylor cone angle value of $49.3^\textrm{o}$, except perhaps at the very moment of blowup.

Different interface tracking schemes have since been implemented by other groups interesting in similar systems. For example, Garzon, Gray and Sethian \cite{Garzon:pre2014} have examined the behavior of perfectly conducting, inviscid droplets situated in an initial uniform external electric field. A level set method was used to develop an Eulerian potential flow model for tracking drop evolution past breakup) while the flow velocity, Maxwell and capillary pressure along the free surface were extracted from axisymmetric boundary integral simulations. Depending on the value of the initial field strength, the resulting exponents at the apical tip for the Maxwell pressure, capillary pressure and kinetic energy density were found to range from 0.662 - 0.676, 0.57 - 0.62 (smaller precision reported due to larger fluctuations) and 0.672 - 0.736, respectively. In all cases examined, larger initial field strengths produced conic tips with larger interior half-angles, although the values were always smaller than the Taylor angle of $49.3^\textrm{o}$.

\subsection{Self-Similar Cusp Formation in the Stokes Regime ($\textsf{Re}=0$)}
\label{BeteluFontelos}

Additional studies during the past decade or so have examined the role of viscous forces in the formation of conic cusps. Vantzos and co-workers \cite{Betelu:pof2006,Fontelos:pof2008} employed boundary integral techniques for tracking the distortion of perfectly conducting droplets to quantify viscous effects. Their studies, focused exclusively on the Stokes flow regime ($\textsf{Re} = 0$), have shown that for sufficiently large surface charge and/or sufficiently large external field, the polar caps of an initially rounded droplet deform into conic tips through a self-similar process as well. The interior half-angles of conic tips that form in charged or uncharged droplets range from about $25^o$ to $30^o$, much smaller than the Taylor angle value of $49.3^\textrm{o}$, irrespective of the values of the actual surface charge or field strength. These studies indicate that viscous effects tend to repress fluid tips from undergoing excessive sharpening.

In an initial study, Betel$\acute{u}$ \textit{et al}. \cite{Betelu:pof2006} examined theoretically and numerically the deformation of an isolated charged droplet of a perfectly conducting viscous liquid surrounded by a dielectric viscous medium in which the  only acting forces were electrostatic repulsion and capillarity. Since all the electric charge resides at the liquid/liquid interface, the surface charge density is given by $\sigma = - \epsilon_o \partial V/\partial n$, where $V$ is the electric potential and $\partial /\partial n$ denotes the normal derivative. The repulsive electrical force per unit area is therefore given by $\bm{F} = \sigma\bm{\hat{E}}/2 = (\epsilon_o/2)(\partial V/\partial n)^2\, \bm{n} = (\sigma^2/2\epsilon_o)\,\bm{n}$. The interior ($i=1$) and exterior ($i=2$) liquid domains satisfy the Stokes equation and continuity equation given by
\begin{equation}
\label{eqn:stokes1}
\mu_{i}\nabla^2\bm{u}_{i}-\nabla p_{i}=0
\end{equation}
\begin{equation}
\label{eqn:stokes2}
\nabla\cdot\bm{u}_{i}=0.
\end{equation}
At the moving interface, the velocity and pressure fields must  satisfy the normal stress boundary condition
\begin{equation}
\label{eqn:NormalBC}
p_{1}- p_{2}=\gamma \kappa + \frac{\epsilon_o}{2}
(\bm{\hat{E_2} \cdot n})^2 + \bm{n} \cdot \left(\bm{\tau}_{1}-\bm{\tau}_{2}\right)\cdot \bm{n}
\end{equation}
and the shear-free boundary condition
\begin{equation}
\label{eqn:TangBC}
\bm{t} \cdot \left(\bm{\tau}_{2}-\bm{\tau}_{1}\right) \cdot \bm{n} =0 \, ,
\end{equation}
where $\kappa=\nabla \cdot \bm{n}$ is the local mean interfacial curvature, $\gamma$ is the interfacial tension and $\bm{\tau}_{i}$ denotes the deviatoric stress tensor for a Newtonian liquid given by
\begin{equation}
\label{eqn:FonStressTensor}
\bm{\tau}_{i}=\mu_{i}\left[\nabla \bm{u}_{i}+(\nabla \bm{u}_{i})^T\right] \,.
\end{equation}
Here, $T$ indicates the matrix transpose, $\bm{n}$ denotes the local unit normal vector pointing outward from medium 1 to 2, which is perpendicular to the two surface unit tangent vectors denoted by $\bm{t}$. The numerical results and supporting scaling arguments show that for isolated charged droplets, the interface curvature, charge density, electric field and velocity at the apical tip diverge  according to $\tau^{-1/2}$ as $\tau \rightarrow 0$ where $\tau$ represents the interval of time measured back from the collapse time.
In a second study \cite{Fontelos:pof2008}, they examined charged droplets in an external electric field. For self-similarity to hold, it was found that the surface change density must scale as $\tau^{-1/2}$, the tip curvature as $\tau^{-\alpha}$ and the tip velocity as $\tau^{\alpha -1}$, where $\alpha >0 $ represents a given function dependent on the far field value of the cone interior half-angle. These studies revealed that the external field acting on a charged droplet represents only a vanishingly small perturbation to the dominant divergent field at the apical tip caused by the accumulation of local surface charge in that region. It was for this reason that the interior half-angle of charged droplets was found to be essentially independent of the value of the external field.

More importantly, Fontelos \textit{et al.} realized from these studies that in the limit of vanishingly small $\textsf{Re}$, the dominant force balance leading to self-similar evolution of the conic tip is not due to opposing Maxwell and capillary stresses but opposing Maxwell and viscous stresses. From the leading order balance between viscous and Maxwell forces in Eq. (\ref{eqn:NormalBC}), they concluded that the tip electric field strength scales as $\tau^{-1/2}$, which in turn requires that the tip curvature scale as $\tau^{-\alpha}$ and the tip velocity scale as $\tau^{\alpha - 1}$,  where $\bm{\xi} = \bm{r}/\tau^{\alpha}$ denotes the self-similar coordinate.

In a separate calculation, they solved for the velocity and electric field distributions satisfying the Stokes and Laplace equation within an infinite cone of perfectly conducting fluid for variable interior half-angle. The velocity field yielded two solutions - one describing  symmetric formation of cusps at opposite poles of an isolated charged droplet or an uncharged droplet in an external electric field and the other describing asymmetric formation of a single cusp.  In both cases, the velocity field, found to scale as $\xi^{(\alpha - 1)/\alpha}$ (derivation not shown), was required to smoothly and asymptotically match the far field velocity function obtained for an equipotential liquid cone subject to vanishing surface shear stress and normal stresses, which scales as $r^{\lambda}$, yielding the important relation $(\alpha - 1)/\alpha = \lambda$.  In the symmetric case, the scaling parameter $\lambda$ was found to be constrained to the value $\lambda=-1$, yielding $\alpha=1/2$.  In the asymmetric case, $\lambda$ was found to be a function of the interior cone angle, in which case neither $\lambda$ nor $\alpha$ yield simple integer ratios. Their simulations produced relatively constant values of the cone interior half-angle of about $27.5^o$ for a wide range of external fields, corresponding roughly to a value $\alpha=-0.74$, in excellent agreement with the exponent value $0.723$ for the tip curvature extracted from the simulations.

\section{Computational Model for Perfectly Conducting Electrified Viscous Liquid in Vacuum}
\label{Formulation}

As discussed above, both the inviscid and Stokes flow regimes permit derivation of the exponents characterizing self-similar growth of the apical tip in perfectly conducting droplets and films. The exponents corresponding to the Maxwell pressure, capillary pressure, kinetic energy density and viscous stresses on approach to the collapse point indicate that for inviscid flow ($\textsf{Re} \rightarrow \infty$), the dominant opposing stresses are given by the Maxwell pressure which promotes tip sharpening and the capillary pressure which represses formation of strong curvature. By contrast, the dominant opposing stresses in the Stokes flow regime ($\textsf{Re} = 0$) have been identified as the Maxwell pressure and the viscous normal stress. In our investigation which follows, we have explored whether  self-similar evolution of the electrified tip persists for all Reynolds numbers, how the interior cone angle depends on the  Reynolds number $\textsf{Re}$ at fixed capillary number $\textsf{Ca}$, and extraction of exponent values characterizing blow up behavior for the various forces acting at the tip.

\begin{figure}
\includegraphics[width=8.5cm]{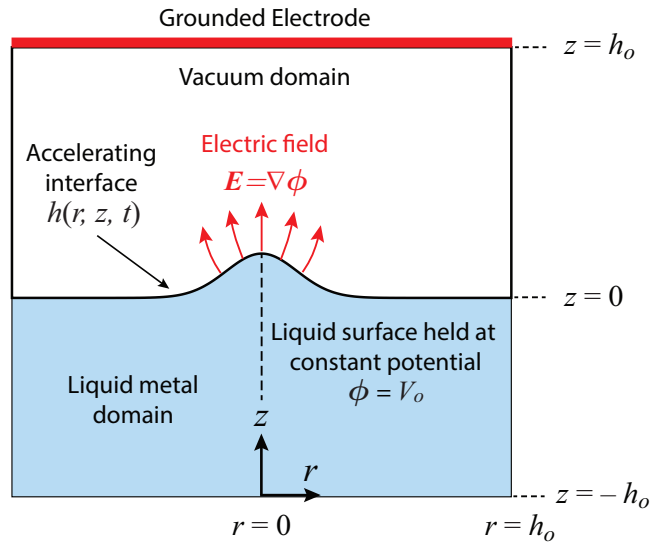}
\caption{Geometry for the computational study. An electrified perfectly conducting layer of liquid metal of radius and thickness $h_o$ held at constant potential $\phi=V_o$ is separated from a grounded circular counter electrode by a vacuum layer of thickness $h_o$. The initial flat liquid layer has a small Gaussian bump centered about the origin which rapidly accelerates toward the grounded electrode due to the increasing field strength at the sharpening tip.}
\label{fig:SimGeometry}
\end{figure}

Shown in Fig. \ref{fig:SimGeometry} is a sketch of the axisymmetric geometry modelled in this study. The system consists of a perfectly conducting body of liquid of radius and initial thickness $h_o$ held at constant potential $V_o$. The initial liquid layer is separated from a grounded cylindrical electrode of radius $h_o$ by a vacuum layer also of thickness $h_o$. This is the same geometry in the original study of Suvorov and Litvinov \cite{Suvorov:jpd2000} and Suvorov and Zubarev \cite{Suvorov:jpd2004} although those simulations were constrained to two values of the Reynolds number, namely $\textsf{Re} = 717$ and $\textsf{Re} = 178$, respectively.

Since charge transport in perfectly conducting liquids is essentially instantaneous in comparison to the time scale for viscous flow, the electrostatic approximation $\nabla \cdot \bm{\hat{E}} =0$ can be used in the vacuum domain where $[0 \le r \leq h_o, h(r,t=0) \leq z \leq h_o]$ where the electric potential $\phi$ then satisfies the equation
\begin{equation}
\nabla^2\phi=0
\end{equation}
subject to the boundary conditions:
\begin{align}
\phi(r,z=h(r,t))&=\phi_o\\
\phi(r,z=h_o,t)&=0\\
\frac{\partial \phi}{\partial r}(r=0,z,t) &= 0\\
\frac{\partial \phi}{\partial r}(r=h_o,z,t) &= 0 \,.\\
\end{align}
Accordingly, the electric field increases in time in response to the advance of the equipotential boundary $h(r,z,t)$ toward the grounded electrode. Any protrusion along the liquid interface therefore undergoes local accelerated growth due to unbalanced field enhancement at the tip.

The hydrodynamic behavior of the electrified liquid is described by the Navier-Stokes equation for an incompressible, Newtonian liquid  where the velocity field $\bm{u}(r,z,t)=u(r,z,t) \bm{\hat e}_r+v(r,z,t)\bm{\hat e}_z$ and pressure field $p(r,z,t)$ satisfy the governing mass and momentum conservation equations given by
\begin{align}
\nabla \cdot \bm{u}&=0\\
\rho\left(\frac{\partial \bm{u}}{\partial t}+(\bm{u}\cdot \nabla)\bm{u}\right)&=-\nabla p+\mu \nabla^2\bm{u} \, .
\end{align}
For simplicity, gravitational effects are neglected by requiring that the Bond number $\textsf{Bo} = \rho g h^2_o/\gamma << 1$. Within the liquid domain $[0 \leq r \leq h_o, \, -h_o \leq z \leq h(r,t)]$, the velocity fields $u(r,z.t)$ and $v(r,z,t)$ must satisfy the following boundary conditions prescribed along the axis of symmetry at $r=0$ and the bottom and side walls:
\begin{align}
u|_{r=0} &=0 & \textrm{symmetry~}\\
\frac{\partial v}{\partial r}\bigg|_{r=0} &=0 & \textrm{symmetry~}\\
u|_{r=h_o} &=0 & \textrm{inpenetrability~}\\
v|_{z=-h_o} &=0 & \textrm{inpenetrability~}\\
u|_{z=-h_o} &=0 & \textrm{no slip~}\\
v|_{r= h_o} &=0 & \textrm{no slip}.
\end{align}
The two remaining boundary conditions for $(u,v)$ are obtained from the constraint that the moving interface $z=h(r,z,t)$ is shear-free, according to which
\begin{equation}
\label{eqn:tangstress}
\bm{n}\cdot\bm{\tau}\cdot\bm{\bm{t}}\big|_{z=h(r,t)} = 0 \, ,
\end{equation}
where $\bm\tau$ is the deviatoric stress tensor defined in Eq. (\ref{eqn:FonStressTensor}). The boundary condition for the pressure field is enforced by the interface normal stress boundary condition
\begin{equation}
\label{eqn:normal1}
p\big|_{z=h(r,t)}=\gamma\left(\nabla\cdot\bm{n}\right)-\frac{\epsilon_o}{2} \hat{E}^2_n + \bm{n}\cdot\bm{\tau}\cdot\bm{n}\, ,
\end{equation}
where $\hat{E}_n = \bm{\hat{E} \cdot n}$. The local mean curvature of the fluid interface $\mathcal{K}$ is defined by the relation $2 \mathcal{K} = - \nabla\cdot\bm{n}$ where in cylindrical coordinates
\begin{align}
2 \mathcal{K} &= b^{-3} \left(\frac{d^2h(r,t)}{dr^2} + \frac{b^2}{r} \frac{dh}{dr}\right)\\
b &=\left[1 + \left(\frac{dh(r,t)}{dr}\right)^2 \right]^{1/2} \,.
\end{align}
The terms on the right hand side of Eq. (\ref{eqn:normal1}) represent the capillary pressure, Maxwell pressure and viscous stress, respectively. The sign convention is such that the capillary pressure at the liquid apex is always positive and therefore repressive due to the concave shape while the Maxwell pressure is always negative, continually pulling the tip toward the counter electrode. The position of the advancing interface is obtained from the kinetic condition which requires
\begin{equation}
v|_{z=h(r,t)} = \frac{\partial h}{\partial t} + \bm{u}|_{z=h(r,t)} \cdot \nabla_s h(r,t)
\end{equation}
where the surface gradient operator is given by $\nabla_s = (\mathbb I - \bm{n}\bm{n}) \cdot \nabla$ where $\mathbb I$ is the identity tensor.

The initial condition for the velocity field represents a quiescent state such that $\bm{u}(r,z,t)=0$. The initial liquid interface configuration represents a flat liquid layer with a small Gaussian bump centered at the origin given by
\begin{equation}
h(0 \leq r \leq h_o,\,t=0)= h_o + a_o e^{-(r/\lambda)^2} \, ,
\end{equation}
where the Gaussian peak amplitude and lateral extent are scale uniformly with the characteristic thickness $h_o$ such that $a_o=0.02 \, h_o$ and $\lambda=0.2 \, h_o$.

\subsection{Scaling and Non-Dimensionalization}

To probe the influence of inertial and viscous forces on the process of cusp formation, we implemented the scalings in Table \ref{table:nd} to convert equations and boundary conditions to dimensionless form. In addition to the parameter values specified in
Fig. \ref{fig:SimGeometry}, the material constants characterizing the liquid metal, namely $\rho$, $\mu$ and $\gamma$, represent the liquid density, viscosity and surface tension, respectively, at the specified temperature.
\begin{table}
\centering
\begin{tabular}{l l l}
\hline
\hline
&& \\
~~~~~Quantity & Scaling & Rescaled \\
&& \,variable\\
\hline
&& \\
\, Coordinates & $r_c = h_o$ & $R = r/r_c$ \,  \\
& $z_c = h_o$  & $Z = z/z_c$ \, \\
&& \\
\, Operators & $r_c = h_o$ & $\widetilde{\nabla} = h_o \nabla$ \,  \\
& $z_c = h_o$  & $\widetilde{\nabla}^2 = h^2_o \nabla^2$ \, \\
&& \\
\, Interface shape & $z_c = h_o$ & $H = h/z_c$ \, \\
&& \\
\, Electric potential & $\phi_c=\phi_o$ & $\Phi=\phi/\phi_c$ \, \\
&& \\
\, Electric field & $E_c=\phi_c/z_c$ & $\bm{E}=\bm{\hat{E}}/E_c$ \, \\
& \,\,\,\,\,\,\,\,$=\phi_o/h_o$ & \, \\
&& \\
\, Pressure & $p_c=\epsilon_o E_c^2/2$ & $P=p/p_c$ \,\\
& \,\,\,\,\,\,\,$=\epsilon_o \phi_o^2/(2h_o^2)$ & \,\\
&& \\
\, Velocity & $u_c = \sqrt{p_c/\rho}$ & $U = u/u_c$ \,\\
& $v_c = \sqrt{p_c/\rho}$ & $V = v/v_c$  \,\\
& \,\,\,\,\,\,$=\sqrt{\epsilon_o/2 \rho}\,(\phi_o/h_o)$&\\
&& \\
\, Time & $t_c = h_o/v_o$ & $T = t/t_c$  \,\\
& \,\,\,\,\,\,$=\sqrt{2\rho/\epsilon_o}\,(h_o^2/\phi_o)$& \,\\
&& \\
\, Self-similar & Fitted exponent & $T_\textsf{C}$\\
\, collapse time\\
&& \\
\, Self-similar && $\tau = T_\textsf{C} - T $\\
\, time interval && \\
&& \\
\, Reynolds & $\textsf{Re}= \rho v_c z_c/\mu$ & \,\\
\, number & \,\,\,\,\,\,\,\,$=\sqrt{\epsilon_o \rho/2 \mu^2} \,\phi_o$ & \,\\
&& \\
\, Capillary & $\textsf{Ca}=p_c z_c/\gamma$& \,\\
\, number& \,\,\,\,\,\,\,\,$=(\epsilon_o /2\gamma)\, (\phi_o^2/h_o)$ & \, \\
&& \\
\, Bond & $\textsf{Bo}=\rho g h^2_o/\gamma$& \,\\
\, number&& \\
&& \\
\hline
\hline
\end{tabular}
\caption{Characteristic scalings (lower case) and non-dimensional variables (uppercase) for system shown in Fig. \ref{fig:SimGeometry}.}
\label{table:nd}
\end{table}

The dimensionless Navier-Stokes equation is then given by
\begin{equation}
\frac{\partial\bm{U}}{\partial T}+\bm{U}\cdot \widetilde\nabla \bm{U}=-\widetilde\nabla P +\frac{1}{\textsf{Re}}\widetilde\nabla^2\bm{U}
\label{eqn:NDNavStokesVec}
\end{equation}
subject to the (dimensionless) normal stress boundary condition at the moving interface:
\begin{equation}
P\big|_{Z=H(R,T)}=\frac{1}{\textsf{Ca}}\left(\widetilde\nabla \cdot \bm{n}\right) - E^2_n + \frac{1}{\textsf{Re}}\bm{n}\cdot \bm{\widetilde \tau} \cdot \bm{n},
\label{eqn:NonDimNormalBC}
\end{equation}
where $E_n = \bm{E \cdot n}$. In rescaled form, the initial interface shape is given by
\begin{equation}
H(R,T=0)=A_o \,e^{-R^2/\Lambda^2}
\end{equation}
where the peak amplitude and extent are given by $A_o=0.02$ and $\Lambda=0.2$. The remaining dimensionless equations and boundary conditions are easily obtained and not restated here.

\subsection{Computational Study of Inertial and Viscous Effects at Fixed \textsf{Ca}}

From Eqs. (\ref{eqn:NDNavStokesVec}) and (\ref{eqn:NonDimNormalBC}), it is evident that both the Reynolds and capillary number play a key role in the formation process. For too small a value of the capillary number, the flow will be overwhelmed by the force of surface tension and budding protrusions will be rapidly leveled. Too small values of $\textsf{Ca}$ will therefore suppress formation of fluid elongations. By contrast, too large a value of $\textsf{Ca}$ will generate  ripples along the moving interface and complicate analysis because of interference effects involving multiple accelerating protrusions.

In this study, we therefore chose to fix the capillary number at a value which would allow the budding and advance of a single protrusion, as in the previous numerical studies of Suvorov \cite{Suvorov:surf2004} and Suvorov and Zubarev \cite{Suvorov:jpd2004}. To that end, we note that in their studies they chose to model the liquid metal gallium at the melting point with $\gamma=0.72$ N/m and a layer thickness $h_o= 5~\mu \textrm{m}$ maintained at a voltage potential $\phi_o$ = 2.4 kV. From the scalings given in Table \ref{table:nd}, the corresponding capillary number in their simulations was $\textsf{Ca}=7.0834$. We adopted this same value for all simulations presented in this work and indeed confirm that this value allowed stable formation of a single conic cusp unaccompanied by any interfacial oscillatory effects.

\subsubsection{Consequences of holding \textsf{Ca} constant in simulations}

In the results which follow, it is important to understand the consequences of holding the capillary number fixed when examining the flow behavior and interfacial shapes obtained at different Reynolds numbers. In particular, given a liquid with fixed material constants, the restriction of fixed capillary number imposes the key constraint that the ratio $\phi^2/ h_o$ is a constant. Therefore, an increase in $\textsf{Re}$ by a factor $\textsf{f}$ reflects an increase in the voltage potential to $\textsf{f} \phi_o$ and an increase in the characteristic length scale $h_o$ to  $\textsf{f}^2 h_o$. According to the scalings in Table \ref{table:nd}), these increases lead to an effective \textit{decrease} in the initial field strength value from $E_o = \phi_o/h_o$ to $E_o/\textsf{f}$ and a corresponding \textit{decrease} in the characteristic flow speeds by a factor $1/\textsf{f}$.

\subsection{Details of Numerical Simulations}
\label{NumericalSimulation}

To solve the coupled equations and boundary conditions discussed in the previous section, we employed finite element simulations using the commercial software package COMSOL \cite{COMSOL} using its  arbitrary Lagrangian-Eulerian method for coupling the vacuum and liquid domains across the moving interface. The computational model was constructed assuming incompressible laminar flow conditions within the liquid domain, modeled as a perfect liquid conductor in contact with an inert vacuum domain. The liquid domain therefore contained no internal electric field and the electric field at the vacuum/liquid interface was always everywhere oriented in the  direction normal and away from the moving interface. The electric potential distribution within the vacuum domain and on the liquid surface was updated at each time step according to the local position of the moving boundary modeled by a moving mesh. At each time step, all mesh element edges along the vacuum/liquid boundary were translated a distance given by the local speed of the liquid. All interior mesh elements were reconfigured by a uniform Laplace smoothing process so that boundary displacement and subsequent fluid motion proceeded smoothly throughout the liquid domain. The entire mesh was reconstituted whenever the mesh quality of the most distorted element fell below a specified value. Mesh generation relied on triangular quadratic Lagrangian elements for the velocity field and linear elements for the pressure field, so-called $P2 + P1$ mixed order discretization. Second-order elements provided higher resolution with stability even at the highest Reynolds numbers.

Two different non-uniform meshes were employed, one optimized for low $\textsf{Re}$ (mesh A) and the other optimized for high $\textsf{Re}$ (mesh B). Flow at intermediate values of $\textsf{Re}$ yielded similar results used either meshing scheme. For mesh A, the edge length of mesh elements adjacent to the moving interface were set equal to the value $0.004 r_o$, where $r_o$ denotes the local interface radius of curvature, subject to an upper and lower bound. Additionally, at very late times where the curvature of the advancing cone increased rapidly, the mesh in the vicinity of the conic tip  was iteratively refined by further dividing the triangular elements to achieve satisfactory resolution. To enforce mesh refinement, the  position of the final location of the conic tip was first estimated from simulations conducted with a less refined mesh. The run was then  repeated with significantly more mesh elements placed near the tip. The total number of mesh elements for the initial and final states ranged approximately from $6000$ to $200,000$. For mesh B, the number of elements positioned adjacent to the interface was directly prescribed, as was the ratio of the edge length of the smallest element adjacent to the conic tip to that of the largest element adjacent to the outermost domain boundary. The mesh was then iteratively refined by splitting elements in the vicinity of the tip, as done with mesh A.  The number of elements for mesh B for the initial and final states ranged from approximately $30,000$ to $70,000$.

Time integration was conducted using a variable first or second order backwards differentiation scheme with adaptive time stepping for a total of about 1000 time steps per simulation.  Simulations were allowed to run just until interface acceleration near the tip led to irreparable mesh inversion. In all cases, the results reported in this work reflect simulations that were all terminated when the capillary pressure at the conic tip reached a fixed imposed value. This was enforced by stopping runs when the dimensionless radius of curvature at the conic tip (scaled by $h_o$) attained the value $2.722 \times 10^{-4}$. In what follows, the final time at which this point was achieved is denoted by $T_f$. This cutoff length scale typically corresponded to a length ten times smaller than the minimum mesh edge length along the moving interface. In previous work modeled on specific liquid metals \cite{Suvorov:surf2004,Suvorov:jpd2004}, researchers have instead terminated simulations once the maximum electric field strength at the conic tip exceeded the field emission value.

\section{Results}
\label{Results}

The following simulations represent results obtained for Reynolds number ranging from $0.1 \leq \textsf{Re} \leq 5 \times 10^4$ at fixed capillary number \textsf{Ca}=7.0834. Details of the evolving  interface shape and exponents characterizing the divergent behavior of governing forces at the conic tip are discussed next.

\subsection{Sharpening of conic cusp}
\begin{figure*}
\includegraphics[width=17.0cm]{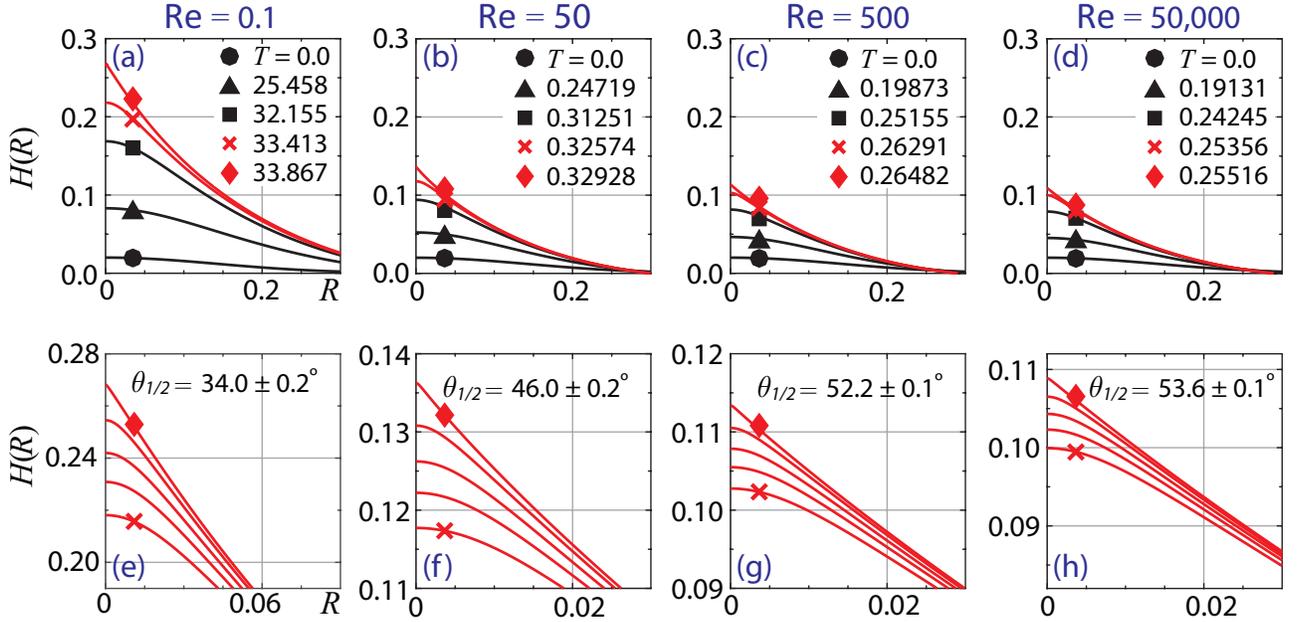}
\caption{Far field (upper panel) and magnified (lower panel) images  (aspect ratio 1:1) of dimensionless interface shapes $H(R,T)$ at times $T$ shown for $Re = 0.1, 50, 500, 50,000$ at $\textsf{Ca}=7.0834$. (a) - (d): Snapshots of axisymmetric interface shapes for $0 \leq T \leq T_f$ where $T_f$ denotes the simulation termination time when the dimensionless curvature radius at the conic tip reached the value $2.722 \times 10^{-4}$. The two uppermost curves (red) bracket the time interval undergoing self-similar growth. (e) - (h): Magnified views of the conic tip during the period of self-similar growth indicated by the red curves in (a) - (d). Time stamps for all intermediate curves are as follows: \textsf{Re}=0.1, $T = 33.593, ~33.713, ~33.809$; \textsf{Re}=50, $T = 0.32710, ~0.32804, ~0.32882$; \textsf{Re}=500, $T=0.26364, ~0.26415, ~0.26457$; \textsf{Re}=50,000, $T=0.25418, ~0.25460, ~0.25495$. Parameter $\theta_{1/2}$ represents the interior half-angle value at $T_f$ obtained from linear regression over the interval $0.005 \leq R \leq 0.030$.}
\label{fig:InterfaceShapes}
\end{figure*}

Shown in Fig. \ref{fig:InterfaceShapes} (a) - (d)  are far field views of the dimensionless interface shape $H(R,T)$ for values of  $\textsf{Re}$ spanning the viscous to inertial regime. For each value of $\textsf{Re}$ shown, the two curves depicted in red in (a) - (d) bracket the time interval observed to undergo self-similar growth, to be discussed in detail below. The last curve for each time sequence shown represents the interface shape at the final simulation time $T_f$, determined from the condition when the dimensionless radius of curvature at the conic tip (scaled by $h_o$) reached the value $2.722 \times 10^{-4}$. Due to viscous retardation effects, it is evident that the time interval required for the apex to attain this small radius of curvature at $\textsf{Re}=0.1$ is approximately two orders of magnitude longer than the time required at higher Reynolds numbers. The peak amplitudes in Fig. \ref{fig:InterfaceShapes} also decrease in magnitude as $\textsf{Re}$ is increased. This can be understood from the fact that at fixed $\textsf{Ca}$, an increase in $\textsf{Re}$ by a factor $\textsf{f}$ leads to a \textit{decrease} in the initial electric field strength to $E_c/\textsf{f}$, which weakens the strength of the Maxwell forces pulling on the liquid tip.

Shown in Fig. \ref{fig:InterfaceShapes} (e) - (h) are magnified  images of the evolving tip during the period of self-similar growth along with the values of the interior half-angle $\theta_{1/2}$ extracted. These angles were obtained from linear regression of the  final interface shape $H(R,T_f)$ over the range $0.005 \leq R \leq 0.030$. (Since $H$ and $R$ are both scaled by $h_o$, this angular value is a constant whether viewed in dimensional or non-dimensional coordinates.) These simulations confirm very rapid formation of conical tips with interior half-angles both above and below the conventional stationary Taylor cone value of $49.3^\textrm{o}$. Larger values of $\textsf{Re}$ are characterized by larger values of $\theta_{1/2}$. This can be understood from the fact that at fixed capillary number, an increase in Reynolds number from $\textsf{Re}$ to $\textsf{f} \textsf{Re}$ leads to a \textit{decrease} in the initial electric field strength from $E_c$ to $E_c/\textsf{f}$. Weaker Maxwell forces lead to formation of wider cones, as expected.

\subsection{Velocity and strain rate profiles near conic apex}
\begin{figure}
\includegraphics[width=8.5cm]{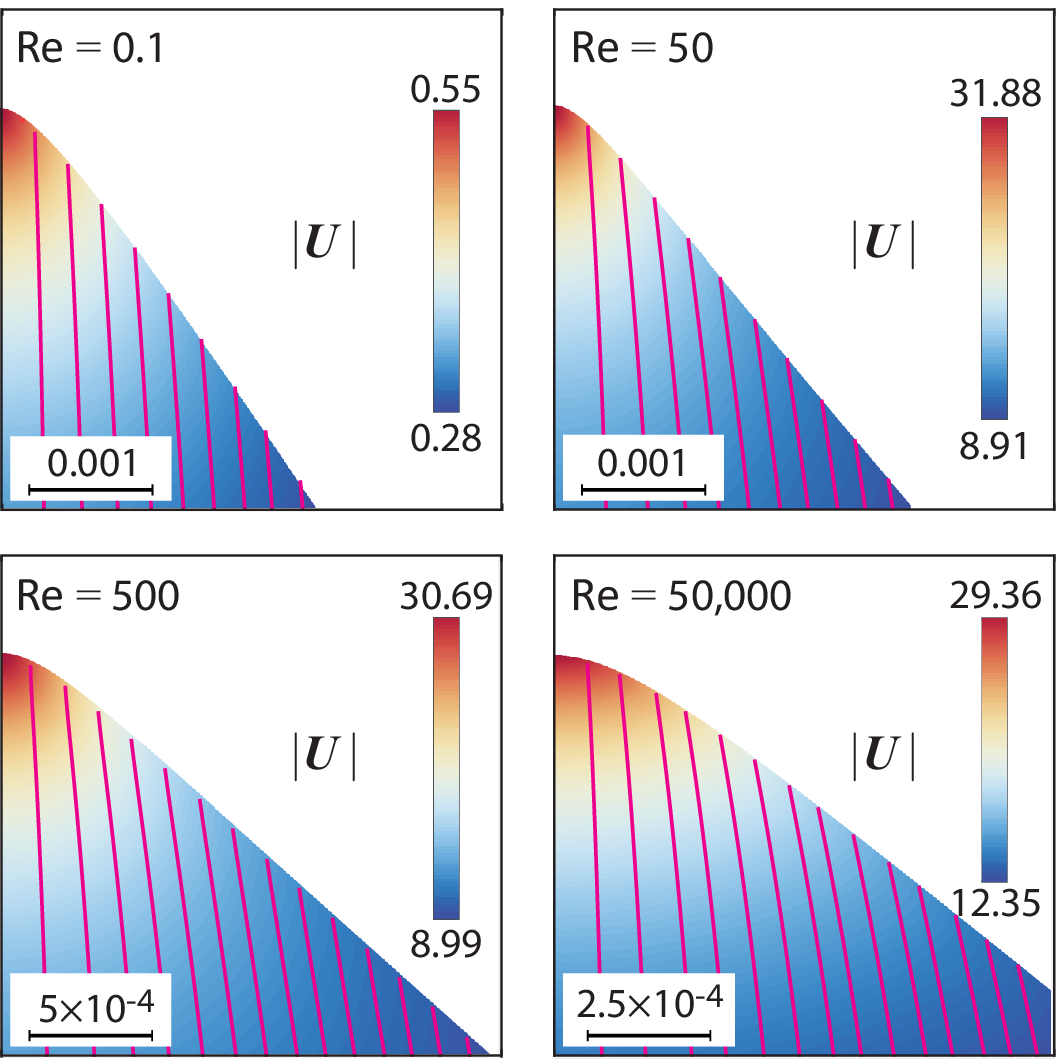}
\caption{Color map of the liquid tip showing the magnitude of the  dimensionless velocity field distribution at the final simulation time $T_f$ for $\textsf{Re}$ = 0.1, 50, 500, 50,000 at $\textsf{Ca}=7.0834$. Scale bars for the dimensionless length scale are shown on the lower left side of each image. The color panes span the minimum and maximum flow speed values indicated below and above the color strip. The curves (magenta) superimposed on the liquid domain represent velocity streamlines.}
\label{fig:Velocity}
\end{figure}

Shown in Fig. \ref{fig:Velocity} are closeup images of the liquid tip at the final time $T_f$ for $Re = 0.1, 50, 500, 50,000$ at $\textsf{Ca}=7.0834$. These are color coded by the magnitude of the dimensionless velocity $|\bm{U}|$. Scale bars are marked as shown in the lower left side of each image. The curves  superimposed on the liquid domain represent the velocity streamlines. These indicate that the net flow at the liquid tip for all \textsf{Re} is oriented in the vertical direction, almost parallel to the vertical axis. The color maps indicate that the highest flow speeds are concentrated near the accelerating tip. However, while the maximum velocity undergoes a significant increase from $Re = 0.1$ to $Re=50$, there is relatively little change in the maximum value  beyond that. This is due to the fact that the simulations were all terminated at about the same value of the (dimensionless) capillary pressure, which effectively establishes a cap on the corresponding  maximum flow speed.

\begin{figure}
\includegraphics[width=8.5cm]{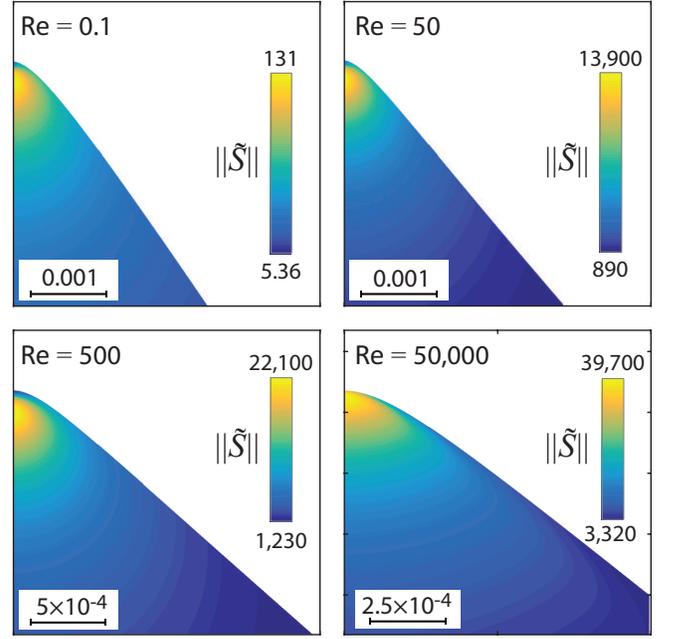}
\caption{Color map of the liquid tip showing the magnitude of the dimensionless shear rate field $\| \widetilde{\bm{S}} \|$ at time $T_f$ for $\textsf{Re}$ = 0.1, 50, 500, 50,000 at $\textsf{Ca}=7.0834$. Scale bars for the dimensionless length scale are provided on the lower left side of each image. The color panes span the minimum and maximum strain rate values indicated below and above the color strip.}
\label{fig:StrainRate}
\end{figure}

Shown in Fig. \ref{fig:StrainRate} are the same set of closeup images as Fig. \ref{fig:Velocity} color coded by the magnitude of the dimensionless rate of strain tensor
$\| \widetilde{\bm{S}} \|= (\widetilde{\bm{S}}:\widetilde{\bm{S}}^{T})^{1/2}/\sqrt{2}$, where $\widetilde{\bm{S}}=[\widetilde{\nabla}\bm{U} + (\widetilde{\nabla} \bm{U})^T]/2$. Scale bars are marked as shown in the lower left side of each image. The region of largest strain rate occurs just below the surface of the conic tip. This region, which sustains the largest rate of deformation, undergoes the highest acceleration. The extent of this region of highest strain rate diminishes in size as the Reynolds number is increased.

\begin{figure}
\includegraphics[width=8.5cm]{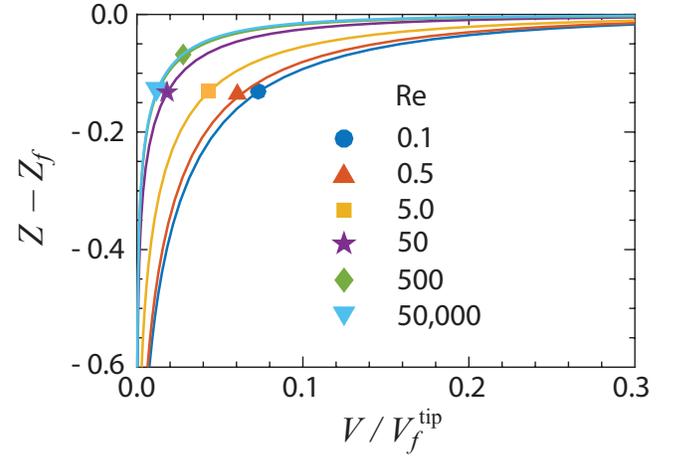}
\caption{Normalized dimensionless velocity $V(R=0,Z,T_f)/V^{\textrm{tip}}_f$ versus distance $Z - Z_f$ for increasing values of $\textsf{Re}$ where $V^{\textrm{tip}}_f=V(R=0,Z_f,T_f)$, $Z_f$ is the final tip vertical location and $T_f$ is the final simulation time.}
\label{fig:BoundaryLayer}
\end{figure}

The very large strain rates observed in the vicinity of the accelerating conic tip give to an interfacial boundary layer. Examination of the vertical velocity component along the axis of symmetry $R=0$ reveals the behavior of this layer with increasing \textsf{Re}. Shown in Fig. \ref{fig:BoundaryLayer} are curves of the normalized dimensionless velocity $V(R=0,Z,T_f)/V^{\textrm{tip}}_f$ versus distance $Z - Z_F$ for increasing values of $\textsf{Re}$ where $V^{\textrm{tip}}_F=V(R=0,Z_f,T_f)$, $T_f$ denotes the final simulation time and $Z_f$ the vertical coordinate of the conic apex at the final time. As $\textsf{Re}$ is increased at fixed capillary number, the boundary layer becomes thinner, as expected. However, the correlation between the boundary layer thickness and $\textsf{Re}$ observed in simple systems with constant surface velocity is not observed here since the interface is not only accelerating rapidly but also accelerating non-uniformally due to the spatially varying electric field strength along the moving interface. Further work is being conducted to elucidate the nature of this interfacial boundary layer.

Due to axisymmetry, the quantity $\| \widetilde S \| (R=0,Z)$ reduces simply to the value $\sqrt{2}\lvert \partial V /\partial Z\rvert (R=0,Z)$. Therefore, the regios of highest strain rate in Fig. \ref{fig:StrainRate} correspond to the uppermost portions of the boundary layer where $\partial V/\partial Z$ is largest. Inspection of Fig. \ref{fig:BoundaryLayer} further reveals that the thickness of the ensuant boundary layers is far larger than the vertical displacements in the corresponding interface shapes in Fig. \ref{fig:StrainRate}. The region of largest strain rate is therefore nestled deep within the boundary layer which extends much further into the bulk liquid.

\subsection{Self-similar evolution of conic tip}
We now turn to evidence of self-similar growth irrespective of Reynolds number for dynamic regimes spanning the viscous to inviscid limit. The evolution of the magnitudes of the various forces acting  at the conic apex point, as evaluated from the normal stress boundary condition given by Eq. (\ref{eqn:NonDimNormalBC}) and the Navier-Stokes equation given by Eq. (\ref{eqn:NDNavStokesVec}), are shown to sustain extended self-similar growth upon approach to the collapse time. Self-similar variable transformations obtained from exponents extracted from the numerical simulations conducted at various \textsf{Re} at fixed \textsf{Ca} confirm that the  accelerating liquid tip undergoes collapse toward a conic shape whose interior angle increases with increasing \textsf{Re}.

\subsubsection{Determination of collapse time $T_\textsf{C}$}

In order to extract accurate power law exponents corresponding to the time at which the forces acting at the conic apex undergo divergence, it is first necessary to extract from the simulations the asymptotic collapse time $T_\textsf{C}$. For each run corresponding to a given value $\textsf{Re}$, this collapse time was determined by examining the growth in the Maxwell pressure at the conic apex as a function of the interval $\tau = T_\textsf{C}-T \rightarrow 0$ plotted on double logarithmic axes and then choosing that value $T_\textsf{C}$ which produced the best linear fit over the final two and one half decades in time. Although this fitting procedure was limited to two and half decades in time, the results below indicate that in the majority of cases, power law growth was observed to occur over longer intervals, as evident in Fig. \ref{fig:SelfSimilarity}.

\subsection{Exponents for self-similar growth of normal stresses at conic apex}
\begin{figure*}
\includegraphics[width=17.0cm]{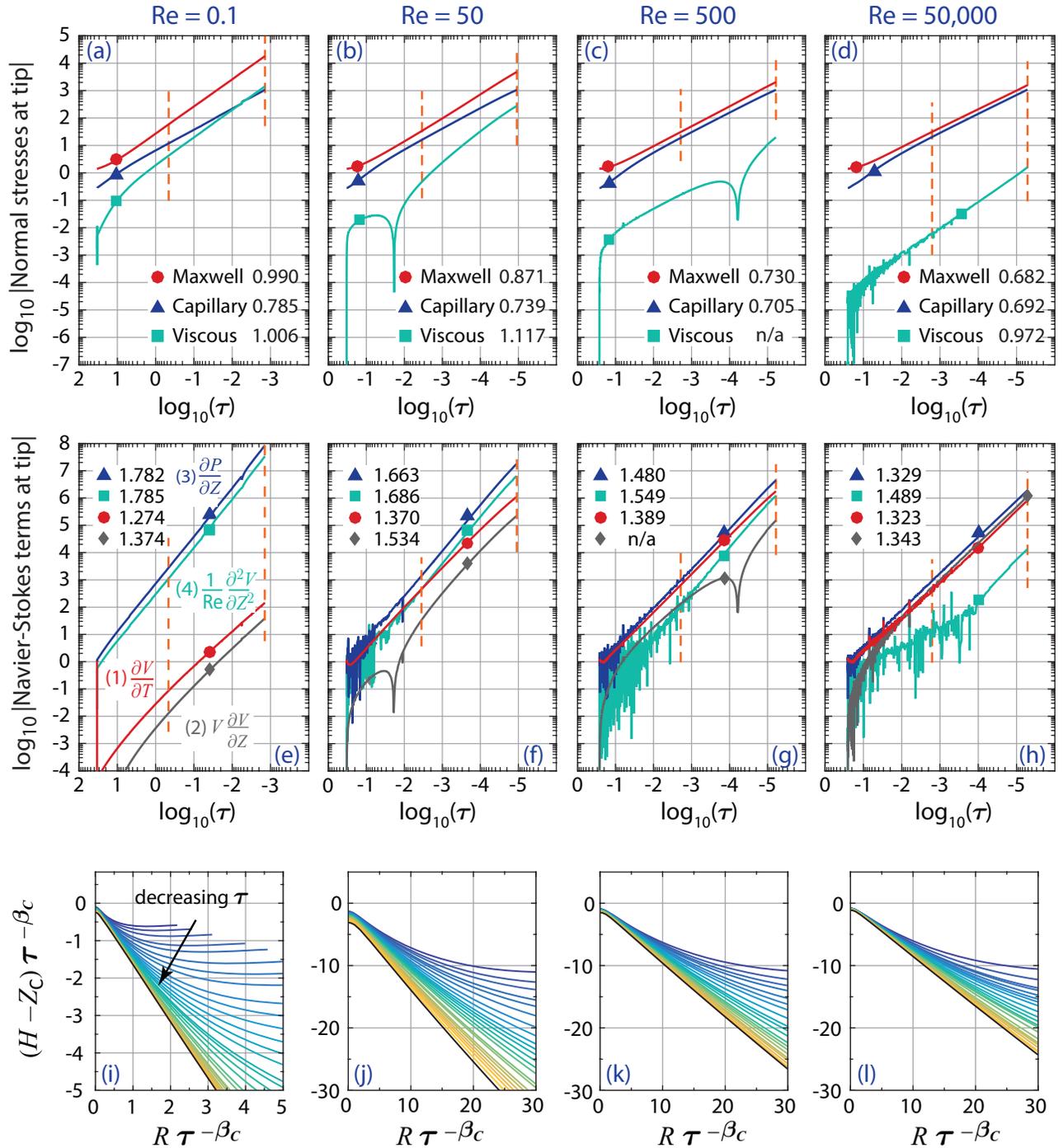}
\caption{Self-similar growth of the forces acting at the conic tip and corresponding extracted exponents. Table \ref{table:betas} contains a complete listing of all exponents extracted. (a)-(d): Double logarithmic plots showing the magnitude of the Maxwell, capillary and viscous normal stresses at the conic apex $(R=0, Z=H_{\textrm{apex}})$ versus $\tau = T_c - T$ computed from Eq. (\ref{eqn:NonDimNormalBC}). Exponent values were obtained by linear regression over the interval marked by the vertical dashed (orange) lines. Exponent values (rounded to 3 digits) appear alongside each normal stress component. (e)-(h): Double logarithmic plots showing the magnitude of the forces per unit volume evaluated at the conic apex versus $\tau = T_c - T$ computed from Eq. (\ref{eqn:NDRzeroNS}). Exponent values were obtained by linear regression over the interval marked by the vertical dashed (orange) lines except for Term 4 (viscous term) where only those data within the final decade in time for $\textsf{Re}>200$ were fitted. Exponent values (rounded to 4 digits) appear next to each icon.  (i)-(l): Self-similar collapse of interface shapes for uniform rescaling of the horizontal and vertical coordinates by  $\tau^{-\beta_C}$ where $\beta_C$ is the capillary stress exponent listed in (a)-(d). The quantity $Z_\textsf{C} \equiv H(R=0,\tau\to 0)$, which denotes the asymptotic (virtual) cone height, was assigned that value which minimized the standard deviation of a linear fit to $H(R=0)-Z_\textsf{C}$ versus $\tau$ on a log-log plot. The quantity $H_\textsf{C}=H(R=0,T=T_\textsf{C})$, which denotes the cone height at the collapse time $T_\textsf{C}$ was assigned that value which minimized the standard deviation of a linear fit to the slope of $H(R,T)$ on a log-log plot for $0.005 \leq R \leq 0.030$. The interface shapes shown in (i)-(l) represent every tenth time step in $\tau$ throughout the period of self-similar growth.}
\label{fig:SelfSimilarity}
\end{figure*}

Shown in Fig. \ref{fig:SelfSimilarity} (a)-(d) are representative double logarithmic plots showing the magnitude of the Maxwell, capillary and viscous normal stresses, reflecting the three terms on the right hand side of Eq. (\ref{eqn:NonDimNormalBC}), for selected values of $\textsf{Re}$ at $\textsf{Ca}=7.0834$. These terms describe the forces per unit area acting at the conic apex. The power law exponents listed alongside each term in (a) - (d) were extracted from those data points corresponding to the final two and half decades in time indicated by the dashed vertical lines, as also marked in images (e) - (h). These results indicate that the Maxwell ($M$), capillary ($C$) and viscous ($V$) normal stresses at the conic apex diverge in time according to $\tau^{-\beta_j}$ where $j = M, C \, \textrm{and}\, V$. Shown in Table \ref{table:betas} are the complete set of exponents and fitting parameters for all simulations conducted in this study. In general, it was found that the exponent values $\beta_M$ and $\beta_C$ tend to decrease with increasing $\textsf{Re}$, indicating that the divergence of these opposing pressures at the conic tip is more rapid at higher Reynolds number.

As evident in Fig. \ref{fig:SelfSimilarity}(a)-(d), the Maxwell stress at the conic apex always exceeds the capillary and viscous normal stress irrespective of Reynolds number. At small values of \textsf{Re}, the Maxwell pressure exceeds the capillary and viscous normal stresses by at least an order of magnitude. As $\textsf{Re}$ increases, the Maxwell and capillary pressures approach each other in  magnitude, while the viscous normal stress decreases in comparison by several orders of magnitude. This trend is understandable since viscous effects always diminish in influence as \textsf{Re} is increased. Furthermore, while at small $\textsf{Re}$ the difference in pressure between the Maxwell and capillary stress increases as $\tau \rightarrow 0$, this difference in pressure is observed to decrease with increasing $\textsf{Re}$ as the collapse time is approached.

With regard to the actual sign of these normal stresses, it is clear from the magnified images in Figs. \ref{fig:Velocity} and \ref{fig:StrainRate} that the liquid shape near the apex is given by a convex curve. The curve representing the capillary stress in Fig. \ref{fig:SelfSimilarity}(a)-(d), computed from the first term on the right hand side of Eq. (\ref{eqn:NonDimNormalBC}), is therefore always positive irrespective of Reynolds number. For the chosen value of \textsf{Ca} in these simulations, the capillary pressure therefore always represses enhancement of tip curvature but is overcome by the Maxwell pressure, computed from the second term on the right hand side of Eq. (\ref{eqn:NonDimNormalBC}), which is always negative and increases in magnitude as the curvature increases. This self-sharpening of the tip goes hand in hand with the local self-enhancement of the electric field in that region. Were it the case that the electric field at the tip is not governed its local value but the global value set by the  externally applied field, then self-similar growth would not be possible since the relative length and time scales could not therefore conform to yield self-similar scalings. This insight is what originally led Zubarev to seek the self-similar exponents reported in Ref. [\onlinecite{Zubarev:jetp2001}] for the inviscid case.

The viscous normal stress, computed from the final term in Eq. (\ref{eqn:NonDimNormalBC}), can assume both positive and negative values depending on $\textsf{Re}$ and $\tau$. For example as shown  in Fig. \ref{fig:SelfSimilarity}(a), the dip in the viscous normal stress at early time near $\log_{10}\tau=-1.6$ is caused by a transition from positive to negative values. Beyond that time, the viscous normal stress becomes increasingly more negative as $\tau \rightarrow 0$. Careful inspection of the velocity gradient $\partial V/\partial Z$ near the tip along the central axis shown in Fig. \ref{fig:StrainRate} confirms that the maximum in the vertical velocity component does not occur at the conic apex but a very short distance below that point. This behavior therefore gives rise to a value $\tau_{ZZ} \propto \partial V/\partial Z < 0$. We are currently exploring whether vorticity effects near the interface are responsible for this small inward displacement of the vertical velocity maximum. Likewise in Fig. \ref{fig:SelfSimilarity}(b), the dip in viscous normal stress near $\log_{10}(T_c-T)=-1.8$ signals a similar transition from positive to negative values. In Figs. \ref{fig:SelfSimilarity}(a) and (b), this transition occurs prior to the onset of the self-similar regime. In Fig. \ref{fig:SelfSimilarity}(c) however, this transition occurs within the self-similar period of growth, which unfortunately precludes extraction of a meaningful exponent - hence the designation ``n/a''. For this reason, we did not report in Table \ref{table:betas} any exponent values $\beta_V$ for $100 \leq \textsf{Re} \leq 1000$. For $\textsf{Re} > 1000$, the viscous normal stress remained positive at all times, allowing extraction of $\beta_V$ from the indicated curve in Fig. \ref{fig:SelfSimilarity}(d) for $\textsf{Re}=50,000$.

\subsection{Exponents for self-similar growth of Navier-Stokes terms at conic apex}

We next evaluate the temporal behavior of each term in the  Navier-Stokes equation at $(R=0, Z=H)$, which due to axisymmetric symmetry, simplifies to the form
\begin{equation}
\underbrace{\frac{\partial V}{\partial T}}_{\textrm{Term 1}}+ \underbrace{V\frac{\partial V}{\partial Z}}_{\textrm{2}}=-\underbrace{\frac{\partial P}{\partial Z}}_{\textrm{3}} +\underbrace{\frac{1}{\textsf{Re}}\frac{\partial^2V}{\partial Z^2}}_{\textrm{4}}\,.
\label{eqn:NDRzeroNS}
\end{equation}
These four terms represent the forces per unit volume acting at that conic apex, which include (1) acceleration, (2) advection, (3) pressure gradient and (4) viscous forces. The double logarithmic plots shown in Fig. \ref{fig:SelfSimilarity} (e)-(h) confirm  self-similar growth of the conic tip as $\tau \rightarrow 0$. As before, we do not report in Table \ref{table:betas} any exponent values $\beta_4$ for the dynamic range $100 \leq \textsf{Re} \leq 1000$ due to the presence of the sharp dip in the viscous stress within the self-similar regime (e.g. Fig. \ref{fig:SelfSimilarity}(g)) caused by a change in sign.

The results in Fig. \ref{fig:SelfSimilarity}(e) confirm that at low Reynolds number such as $\textsf{Re}=0.1$, the dominant competition  is given by the pressure gradient and the viscous term, both roughly five orders of magnitude larger than the inertial terms given by Terms 1 and 2. This difference in magnitude between Terms 3-4 and 1-2 is observed to increase as $\tau \rightarrow 0$. As $\textsf{Re}$ increases, however, the inertial Terms 1 and 2 become more prominent, eventually approaching similar magnitude as the pressure gradient. By contrast, the viscous term decreases in magnitude, becoming rather negligible by about $\textsf{Re} = 20,000 - 30,000$ (not shown). For $\textsf{Re}=50,000$, it is evident from Fig. \ref{fig:SelfSimilarity}(h) that the viscous normal stress at the conic tip is at least two orders of magnitude smaller than the other three terms in the Navier-Stokes equation. In all the simulations conducted, the magnitude of the pressure gradient given by Term 3 exceeded that of the other terms throughout the period governed by self-similar growth. This likely reflects the fact that the rapidly increasing Maxwell pressure at the conic apex remains the dominant component in the pressure gradient irrespective of Reynolds number.

For completeness, we note here the changes in sign which occur in  time for Terms 1 - 4 in Eq. (\ref{eqn:NDRzeroNS}) shown in Fig. \ref{fig:SelfSimilarity}(e)-(h). Term 1, which tracks the acceleration of the conic apex, is always positive irrespective of the value of $\textsf{Re}$. This must be the case since the dominant driving force at the tip is the Maxwell pressure which increases rapidly in time due to the increase in local electric field strength as the tip curvature increases in time. Term 2 tracks the inertial advective term whose sign depends on $\textsf{Re}$ and $\tau$. For $\textsf{Re} = 0.01$, the actual sign of Term 2 is negative, which implies that $\partial V/\partial Z < 0$ at the apex. For $\textsf{Re}$ = 50 and 500, there occurs a sharp dip in Term 2, which signals a transition from positive to negative values as $\tau$ decreases. For $\textsf{Re}=50,000$, there initially occurs significant scatter in Term 2 which eventually settles down to reveal steady power law growth. During this latter period, Term 2 is always positive in value.

Except for some scatter at early times, Term 3 given by the pressure gradient $\partial P/\partial Z$ is always negative irrespective of the Reynolds number, which indicates that the driving pressure gradient $\partial P/\partial Z > 0$. This behavior is expected since the dominant driving force due to the Maxwell pressure at the conic apex always pulls fluid toward the counter electrode. The viscous force per unit volume given by Term 4 is well behaved at small Reynolds number and always negative, counterbalancing the pressure gradient term. The data for Term 4 become significantly noisier with increasing $\textsf{Re}$ while its magnitude decreases significantly, especially within the period of self-similar growth. This behavior can be traced to the rapid change in the vertical velocity component along the central axis within a distance of a few mesh elements of the accelerating interface. For $\textsf{Re}=50$ and 500, this noisy regime occurs prior to the time interval used to extract the corresponding exponent. For $\textsf{Re}=50,000$, this noisy behavior persists well into the self-similar regime. For this reason, the exponents reported for Term 4 for $\textsf{Re}>200$ were extracted using only data generated during the final decade in time where there occurs smooth power law growth.

\subsection{Self-similar collapse of conic tip shape}
The time sequence images in Fig. \ref{fig:SelfSimilarity} (i-l) confirm self-similar collapse of interface shapes obtained by rescaling the horizontal and vertical coordinates by the factor $\tau^{-\beta_C}$, where $\beta_C$ is the exponent for the capillary pressure growth plotted in Fig. \ref{fig:SelfSimilarity} (a)-(d). This dynamic length was chosen to be the rescaling length since it is a direct measure of the diminishing radius of curvature at the conic apex. The curves in Fig. \ref{fig:SelfSimilarity} (i-l) depict the evolving interface shapes during the period of self-similar growth identified in (a) - (h). The non-uniform spacing between curves reflects the adaptive time stepping scheme used in the simulations, as discussed in Section \ref{NumericalSimulation}.

\begin{figure}
\includegraphics[width=8.5cm]{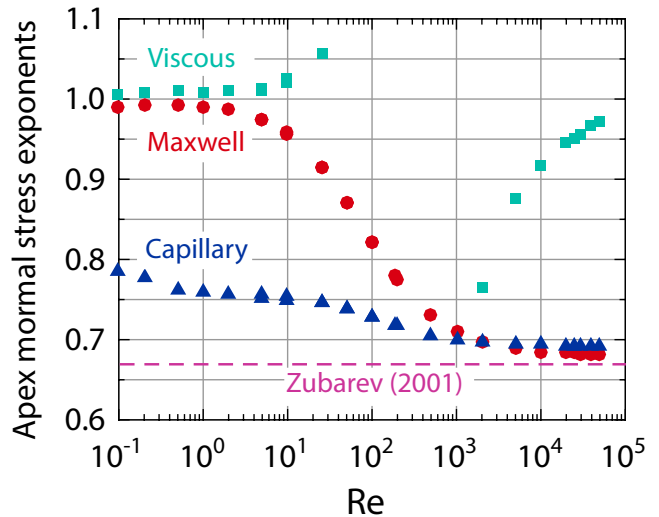}
\caption{Semilogarithmic plot of the magnitude of the exponent values $\beta_M$, $\beta_C$, and $\beta_V$ versus $\textsf{Re}$ computed from the terms in the normal stress boundary condition given by Eq. (\ref{eqn:NonDimNormalBC}). These terms each diverge as $\tau^{-\beta_j}$ where $j=M, C \,\textrm{and}\,V$. The dashed horizontal line shown (purple) indicates the exponent value $\beta_M = \beta_C = 2/3$ derived by Zubarev \cite{Zubarev:jetp2001} for the inviscid limit as $\textsf{Re} \rightarrow \infty$. The complete set of exponent values are provided in Table \ref{table:betas}.}
\label{fig:betasvsReE}
\end{figure}

\subsection{Normal stress exponents at conic apex vs \textsf{Re}}
The simulations conducted in this study, whose range in Reynolds number spans the viscous to the inertial regime, confirm self-similar growth of the conic tip irrespective of Reynolds number. As clear from Fig. \ref{fig:SelfSimilarity} (a) - (h), which pairs of forces predominantly contribute to the self-similar blow up behavior changes as \textsf{Re} increases. Figure \ref{fig:betasvsReE} illustrates the variation in the magnitude of the blow up exponents $\beta_M$, $\beta_C$ and $\beta_V$ for increasing $\textsf{Re}$ at $\textsf{Ca}=7.0834$ computed from the normal stress boundary condition given by Eq. (\ref{eqn:NonDimNormalBC}). The dashed horizontal line represents the exponent values $\beta_M = \beta_C = 2/3$ first predicted by Zubarev \cite{Zubarev:jetp2001} for cone formation in the inviscid limit as $\textsf{Re} \rightarrow \infty$. Error bars, which were always smaller than $10^{-3}$, are not shown since they would appear smaller than the icons used to designate the mean value. Included in Fig. \ref{fig:betasvsReE} are the results for $\textsf{Re}=$ 5, 10, 25, 50 and 100 obtained from the two meshing schemes (A and B) discussed in Section \ref{NumericalSimulation}. These data are nearly indistinguishable, confirming that the two meshing schemes produced virtually identical results.

Fig. \ref{fig:betasvsReE} illustrates which normal stresses at the moving interface maintain similar exponents. At the smallest values of $\textsf{Re}$, the Maxwell and viscous normal stresses (i.e. forces per unit area) assume a value close to one, although as indicated in Fig. \ref{fig:SelfSimilarity}(a), their actual magnitudes differ by more than an order of magnitude. As perhaps more evident from Fig. \ref{fig:SelfSimilarity}(e) which shows the resulting forces per unit volume acting at the tip, the Maxwell and viscous forces do provide the main competitive balance within the viscous dominated regime, while the capillary force per unit volume acts more as a smaller corrective term. By contrast, at much higher values of the Reynolds number, the dominant competition switches to that between the Maxwell and capillary stress while the viscous normal stress appears as a smaller correction. These results confirm the dominant pair forces reported by Fontelos \textit{et al.} in the Stokes limit \cite{Fontelos:pof2008} and Zubarev \cite{Zubarev:jetp2001} in the inviscid limit.

\subsection{Comparison with exponents predicted by Zubarev and Fontelos \textit{et al}.}

The results in Fig. \ref{fig:betasvsReE} for the exponents $\beta_M$ and $\beta_C$ indicate that as the Reynolds number increases toward the inviscid limit, both asymptote toward the value 2/3 first predicted by Zubarev \cite{Zubarev:jetp2001}. Zubarev's scalings in Eq. (\ref{eqn:StartSelfSim})-(\ref{eqn:EndSelfSim}) can also be used to estimate the corresponding asymptotic exponent for the viscous normal stress. Noting that the viscous normal stress in Eq. (\ref{eqn:NonDimNormalBC}) at the apex is given by $\partial V/\partial Z$, it must therefore scale as  $\tau^{-\beta_V}=\tau^{-1}$. Our simulations confirm that at  $\textsf{Re} = 50,000$, the magnitude of this exponent has already reached a value $0.9719 \pm 0.0025$, close to the asymptotic limit of one. The trend evident from Fig. \ref{fig:betasvsReE} is that the magnitude of $\beta_V$ will further smoothly increase toward unity as $\textsf{Re} \rightarrow \infty$.

Zubarev's self-similar analysis requires that lengths scale as $\tau^{2/3}$, velocities as $\tau^{-1/3}$ and the capillary and Maxwell pressures as $\tau^{-2/3}$. In concluding the analysis in his original work, he applied these inviscid scalings to the Navier-Stokes equation given by Eq. (\ref{eqn:NDRzeroNS}) to determine what was the resulting scaling for the viscous stress and found that
\begin{align}
\underbrace{\frac{\partial V}{\partial T}}_{1} \sim \underbrace{V \frac{\partial V}{\partial Z}}_{2} \sim \underbrace{\frac{\partial P}{\partial Z}}_{3} &\sim \tau^{-4/3} \\
\underbrace{\frac{1}{\textsf{Re}}\frac{\partial ^2V}{\partial Z^2}}_{4} &\sim \tau^{-5/3}\,.
\end{align}
Inspection of the exponent values in Table \ref{table:betas} for $\textsf{Re} = 50,000$ from our simulations, namely $\beta_1= 1.3228 \pm 0.0028$, $\beta_2 = 1.3430 \pm 0.0025$ and $\beta_3=1.3294 \pm 0.0008$, confirms close agreement with his inviscid prediction of $4/3$. The value $\beta_4 = 1.4893 \pm 0.0069$ for the viscous stress  is somewhat smaller than his asymptotic estimate of $-5/3 = 1.6667$.

It is worth noting here that although Zubarev's findings stemmed from  a self-similar analysis of Bernoulli's equation under the assumption of inviscid and irrotational flow, the same exponents can be derived  from the Euler equation (i.e. the inviscid limit of the Navier-Stokes equation) and the corresponding inviscid limit of the normal stress boundary condition. A back of the envelope calculation leads to the prediction
\begin{equation}
\beta_M=\beta_C=2/3~~~~\textrm{and}~~~~\beta_V=1.
\end{equation}
Apparently then, the irrotational constraint inherent in the Bernoulli analysis is not a requirement for deriving these exponents. Indeed, in all our studies we have observed the presence of vorticity just below the accelerating interface. In a separate work, we are investigating the behavior of this interface induced vorticity, whose presence nonetheless preserves his original predictions for these exponents.

The results of our simulations also show good agreement with the predictions of Fontelos \textit{et al.} \cite{Fontelos:pof2008} in the Stokes flow limit as $\textsf{Re }\rightarrow 0$. As discussed in Section \ref{BeteluFontelos}, Fontelos \textit{et al.} demonstrated that at $\textsf{Re}=0$, the electric field strength at the apex of an electrified liquid drop must diverge as $\tau^{-1/2}$, or equivalently, that the Maxwell pressure must scale as $\tau^{-1}$, which leads to a value $\beta_M=1$. Our simulations at $\textsf{Re}=0.1$ yield exponent values $\beta_M=0.9897 \pm 0.0002$, in excellent agreement with the asymptotic prediction. Their analysis also predicts that the exponent associated with the capillary pressure must vary with the cone interior half-angle $\theta_{1/2}$ such that $\beta_C=\alpha(\theta_{1/2})$. In Ref. [\cite{Fontelos:pof2008}], the authors provide a numerical plot showing the variation in exponent values corresponding to the flow velocity (and electric potential) as a function of increasing cone interior half angle. From that plot and the value of the interior half angle $\theta_{1/2}=34^o$ extracted from our Fig.\ref{fig:InterfaceShapes} (e) for $\textsf{Re}=0.1$, we determine that $\alpha=0.8465$. Our estimated value of $\beta_C=0.7849 \pm 0.0017$ for $\textsf{Re}=0.1$, which from the trend shown in Fig.\ref{fig:betasvsReE} will clearly increase as the Reynolds number is lowered still, is in good agreement with their asymptotic prediction that $\beta_C = 0.8465$ as $\textsf{Re} \rightarrow 0$.

Noting once again that the viscous normal stress in Eq. (\ref{eqn:NonDimNormalBC}) at the apex is given by $\partial V/\partial Z$, it must then be the case from the scalings for $V$ and $Z$ that the viscous normal stress scales as $\tau^{-1}$, the same scaling obtained in the limit of high \textsf{Re}. Our simulations at $\textsf{Re}=0.1$ yield a value $\beta_V =1.0064 \pm 0.0009$, in excellent agreement with this prediction. In fact, the dimensional analysis requires that $\partial V/\partial Z \sim \tau^{-1}$ regardless of Reynolds number. Inspection of the values extracted for $\beta_V$ given in Table \ref{table:betas} indeed confirm that all the reported values are close to one. The largest deviations from unity occur for those intermediate values of Reynolds number, which as we noted earlier, incur a change in sign within the fitting interval and a corresponding sharp dip, as shown in Fig. \ref{fig:SelfSimilarity} (f) - (h). The occurrence of these dips tends to skew the extracted values of $\beta_V$ slightly away from one.

The analysis by Fontelos \textit{et al.} \cite{Fontelos:pof2008} for self-similarity in the Stokes regime also predicts that $Z$ scales in time as $\tau^{\alpha}$ (where $\alpha > 0$), $V$ as $\tau^{\alpha -1}$ and the voltage potential as $\tau^{\alpha - 1/2}$. The latter relation therefore requires that the electric field strength diverge as $\tau^{-1/2}$ and therefore yields that the Maxwell pressure must diverge as $\tau^{-1}$. Substituting these scalings into Eq. (\ref{eqn:NDRzeroNS}) and the value $\alpha=0.8465$ we determined above for the runs conducted at $\textsf{Re}=0.1$ yields
\begin{align}
\frac{\partial V}{\partial T} &\sim \tau^{\alpha-2}\approx \tau^{-1.1535}\\[0.5 em]
V \frac{\partial V}{\partial Z}&\sim \tau^{\alpha-2}\approx \tau^{-1.1535}\\[0.5 em]
\frac{\partial P}{\partial Z} &\sim \tau^{-1-\alpha}\approx \tau^{-1.8465}\\[0.5 em]
\frac{1}{\textsf{Re}}\frac{\partial ^2V}{\partial Z^2} &\sim\tau^{-1-\alpha}\approx \tau^{-1.8465} \,.
\end{align}
Listed in Table \ref{table:betas} are the values of the corresponding exponents extracted from the simulations run at $\textsf{Re} = 0.1$, namely $\beta_1=1.2735 \pm 0.0021$, $\beta_2=1.3739 \pm 0.0033$, $\beta_3=1.7815 \pm 0.0039$ and $\beta_4=1.7849 \pm 0.0028$, which are all consistent with the asymptotic predictions by Fontelos \textit{et al.} for $\textsf{Re} = 0$.

\section{Discussion}
\subsection{Laminar flow conditions}

\begin{figure}
\includegraphics[width=8.5cm]{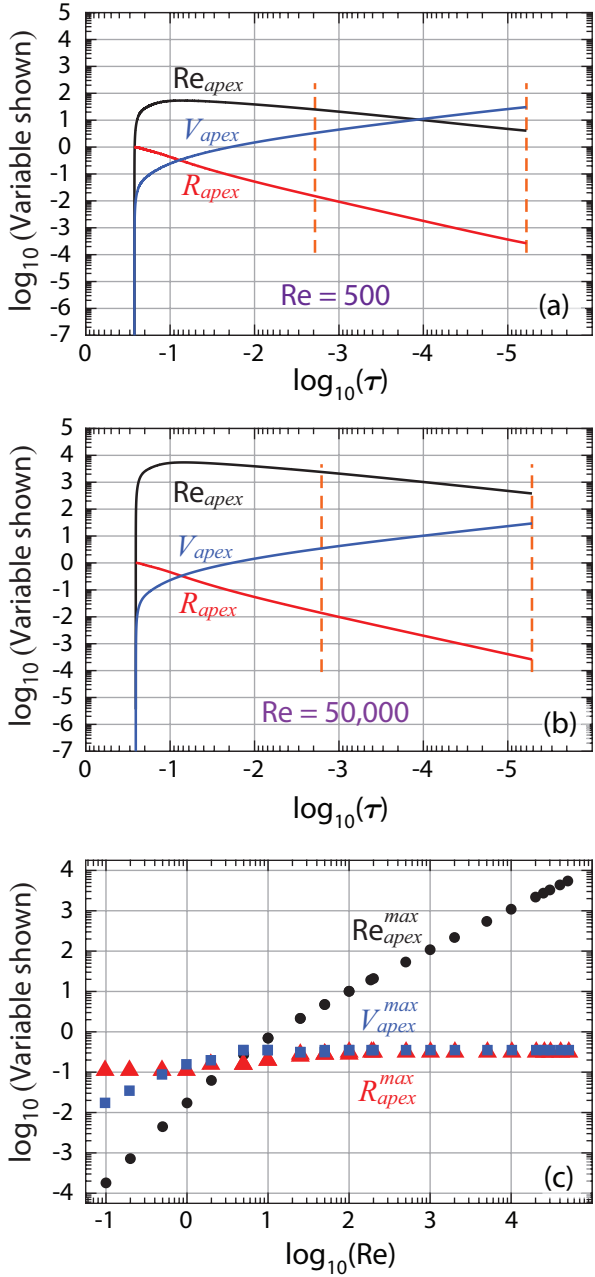}
\caption{Apex curvature, apex flow speed and apex (local) Reynolds number as a function of $\tau=T_C-T$ for $0.1 \leq \textsf{Re} \leq 50,000$ at $\textsf{Ca} = 7.0834$. (a)-(b) Double logarithmic plot showing apex radius of curvature $R_{apex}$, apex velocity $V_{apex}$ apex (local) Reynolds number $\textsf{Re}_{apex}= V_{apex} R_{apex} \times \textsf{Re} $ versus $\tau$ for $\textsf{Re}=500$ and $\textsf{Re}=50,000$ at $\textsf{Ca} = 7.0834$. Dashed vertical lines (orange) correspond to the same interval of self-similar growth shown in Fig. \ref{fig:SelfSimilarity}. (c) Double logarithmic plot showing the maximum apex Reynolds number achieved during each run where $\textsf{Re}^{max}_{apex}=\textsf{Re} V^{max}_{apex}R^{max}_{apex}$. $R^{max}_{apex}$ and $V^{max}_{apex}$ denote the corresponding values of the apex curvature radius and apex vertical speed.}
\label{fig:TipReynolds}
\end{figure}

The parameter range investigated for which $0.1 \leq \textsf{Re} \leq 50,000$ at $\textsf{Ca}=7.0834$ was confirmed to maintain laminar flow conditions throughout. While the Reynolds number defined in Table \ref{table:nd} is useful in cataloguing the general flow behavior discussed, it's value is based on the \textit{initial} electric field strength $\phi_o/h_o$. In order to assess flow conditions in the region closest to the rapidly accelerating tip, it is also useful to define a local value of the Reynolds number given by
\begin{equation}
\textsf{Re}_{apex}=\frac{\rho\,v_{apex}\,r_{apex}}{\mu}\, .
\end{equation}
where $v_{apex}$ denotes the dimensional vertical component of the fluid velocity at the conic apex and $r_{apex}$ the apex radius of curvature. Non-dimensionalizing these local parameters by $v_c$ and $z_c$ defined in Table \ref{table:nd} yields
\begin{equation}
\label{eqn:Retip}
\textsf{Re}_{apex}= V_{apex}\, R_{apex}\times \textsf{Re} .
\end{equation}
Our simulations and analysis indicate that at finite Reynolds number $\textsf{Re}$, the capillary pressure at the apex scales as $\tau^{-\beta_C}$. Since this pressure is inversely proportional to the (dimensionless) apex radius of curvature $R_{apex}$, this implies that $R_{apex} \sim \tau^{\beta_C}$. Furthermore, since the viscous normal stress $\partial V/\partial Z$ at the apex scales as $\tau^{\beta_V}$, then  $V \sim \tau^{\beta_C - \beta_V}$. These scalings therefore suggest that
\begin{equation}
\textsf{Re}_{apex} \sim \tau^{2\beta_C-\beta_V} \,.
\end{equation}
For the exponent values $\beta_C$ and $\beta_V$ listed in Table \ref{table:betas}, it is clear that $2\beta_C-\beta_V >0$ and therefore $\textsf{Re}_{apex} \sim \tau^{2\beta_C-\beta_V}$ decreases as $\tau = T_C - T$ decreases, as confirmed by the curve shown in Fig. \ref{fig:TipReynolds}(a) and (b). In both plots, $R_{apex}$ decreases monotonically as the collapse time is approached due to the sharpening of the conic tip while $V_{apex}$ increases monotonically as the tip continues to accelerate. Although similar to the behavior in Fig. \ref{fig:TipReynolds}(a), the data in  Fig. \ref{fig:TipReynolds}(b) indicate that the values for $\textsf{Re}_{apex}$ increase by approximately two orders of magnitude while $\textsf{Re}$ increases by only two orders of magnitude (from 500 to 50,000). This correlation is clear from the  product $V^{max}_{apex}\,R^{max}_{apex}$ in Fig. \label{eqn:Retip}(c) which remains fairly constant as $Re$ is increased except at the  smallest values of $\textsf{Re}$ in the viscous dominated regime.

We note that in Fig. \ref{fig:TipReynolds}(a), the since the maximum value $Re^{max}_{apex} = 53.68$, which is achieved at $\log_{10}\tau=-1.1441$ (i.e. $T=0.19307$), the flow at the tip maintains laminar conditions. In Fig.\ref{fig:TipReynolds}(b), the maximum value $\textsf{Re}^{max}_{apex} = 5462.4$, which is achieved at $\log_{10}\tau=-1.1535$ (i.e. $T=0.18511$), also confirms laminar flow. The transition to turbulence for external flow typically  sets in at Reynolds numbers of about $2 - 5 \times 10^5$. Interestingly, the maxima in the local Reynolds number in Fig. \ref{fig:TipReynolds}(a) and (b) both establish very early in the liquid deformation process, prior to the onset of self-similar growth.  The early rapid increase in $\textsf{Re}^{max}_{apex}$ followed by a monotonic decrease is due to the fact that the decrease in $R_{apex}$ with time is always more rapid than the corresponding increase in $V_{apex}$.

\subsection{Correlations at conic apex with local Reynolds number}
Shown in Fig. \ref{fig:TipReynolds}(c) is the maximum apex Reynolds number achieved during each run for $0.1 \leq \textsf{Re} \leq 50,000$ at $\textsf{Ca} = 7.0834$ defined by $\textsf{Re}^{max}_{apex}=\textsf{Re} V^{max}_{apex}R^{max}_{apex}$. The quantities $R^{max}_{apex}$ and $V^{max}_{apex}$ denote the corresponding values of the tip curvature radius and tip speed. The function $\textsf{Re}^{max}_{apex}$ increases smoothly and monotonically with increasing \textsf{Re} such that $\textsf{Re}^{max}_{apex} \approx 0.1 \times \textsf{Re}$ for $\textsf{Re}>10$. It is evident from Fig. \ref{fig:TipReynolds}(c) that at intermediate and large values of \textsf{Re}, where the dominant opposing stresses are the Maxwell and capillary stress, the velocity and curvature radius track each other closely in magnitude. This is expected since the Maxwell pressure is always the dominant driving force and its magnitude is set by the apex curvature radius. We note that both $V^{max}_{apex}$ and $R^{max}_{apex}$ are rather independent of  $\textsf{Re}$ (except in the viscous dominated regime at low $\textsf{Re}$). This behavior confirms that these two values are set by local conditions at the apex, where the electric field strength and curvature radius undergo self-enhancement due to local conditions. This stands in contrast to the behavior at very small values of the Reynolds number $\log_{10} (\textsf{Re}) < 1$, where the dominant opposing stresses are the Maxwell and viscous normal stress. The latter introduces longer range interactions mediated by the liquid viscosity which therefore causes a dependence on $\textsf{Re}$. Close inspection of the the curve for $\textsf{Re}^{max}_{apex}$ in Fig. \ref{fig:TipReynolds}(c) reveals a change in slope at a value of about $\log_{10}(\textsf{Re}) = 1$. This break in slope reflects the viscous dominated regime to the left where $\textsf{Re}^{max}_{apex} \sim \textsf{Re}^2$ and the inertia dominated regime to the right where $\textsf{Re}^{max}_{apex} \sim \textsf{Re}$.

\subsection{Measurement restrictions involving liquid film thickness  and electric field strength}

\subsubsection{Restrictions due to Bond number}
In this study, the capillary number was held fixed at a value $\textsf{Ca} = 7.0834$ and gravity effects neglected such that the Bond number $Bo = \rho g h^2_o/\gamma << 1$. The simulations were also terminated once the dimensionless apex radius of curvature (normalized by $h_o$) attained the value $2.722 \times 10^{-4}$. These constraints enforce certain requirements on physical systems that may someday be used to verify the results obtained. Furthermore, it is assumed that electric field values at the apex should never  exceed field ion emission values in liquid metals typically of order $10^{10}$ V/m \cite{Orloff:chapter2008}, though the actual values depend not only on the liquid metal but also the operating temperature and vacuum conditions.

\subsubsection{Restrictions due to maximum field strength}

To ascertain the parameter range of validity for the simulations in this study, we provide in Table \ref{table:materials} a list of derived values for three fluids at their melting temperature commonly  used in liquid metal ion source devices, namely gallium (Ga), indium  (In) and cesium (Cs). The entries were computed as follows. From the scalings in Table \ref{table:nd}, a fixed value of $\textsf{Re}$ for a given liquid system corresponds to a specific initial voltage potential $\phi_o$. The constraint $\textsf{Ca} = 7.0834$ then establishes the specific thickness $h_o$ of the initial liquid layer, which then determines the initial field strength value $E_o=\phi_o/h_o$. The material constants and the value $h_o$ then establish the corresponding value of the Bond number. For ground based experimental studies, it would be necessary to satisfy the inequality $Bo << 1$ when making comparison to the findings of our study. Of course, one can always redo our numerical study using a larger value of $\textsf{Ca}$, which for a given value of $\phi_o$ would ensure a smaller value of $h_o$ and therefore a smaller Bond number. Too large a value of $\textsf{Ca}$, however, can trigger surface waves leading to multiple protrusions and interference effects.

Another important consideration for experimentalists is the small time window for measurement before onset of ion emission, a phenomenon that has not yet been incorporated into numerical models designed to probe conic formation. The difficulty in doing so stems from the very disparate length and time scales characterizing the hydrodynamic flow and field evaporation or related process. The initial field strength value in simulations should therefore not exceed field emission values. Shown in Table \ref{table:materials} are values for the initial electric field strengths $E_o$ for the specified values of $\textsf{Re}$ and $\textsf{Ca}$ used in our simulations as determined from the definitions in Table \ref{table:nd}. The values of $E_o$ for the smallest Reynolds number simulated in our study, namely $\textsf{Re} = 0.1$, would likely exceed the limit for ion emission and therefore preclude measurement; the corresponding values of $h_o$ are also unrealistically small. For the material constants listed, measurements at very low Reynolds numbers are likely not possible even at different values of  $\textsf{Ca}$. According to the expressions for $\textsf{Re}$ and $\textsf{Ca}$ given in Table \ref{table:nd}, a fixed value of $\textsf{Re}$ corresponds to a fixed value of the voltage potential $\phi_o$. Therefore smaller values of $\textsf{Ca}$ correspond to larger values of $h_o$ and ultimately smaller values of the field strength $E_o=\phi_o/h_o$. For $\textsf{Re} = 0.1$, the value $E_o$ would have to be reduced by about two orders of magnitude to forestall ion emission, which could be accomplished by increasing $h_o$ by two orders of magnitude. However, that would still lead to unrealistically small values of $h_o$, in the range of 0.1 nm. It is therefore unlikely that self-similar behavior in the Stokes regime can be tested using the system depicted in Fig. \ref{fig:SimGeometry}.

\section{Recent experimental studies of dynamic cone formation in electrified fluids}
Due to the very rapid flow speeds involved in dynamic cone formation in typical electrospray and liquid metal ion sources, some experimentalists have tried to slow the formation process by using  partially conducting fluids beneath a layer of insulating liquid. This has allowed direct optical tracking and quantification of cuspidal acceleration undergoing self-similar growth. Oddershede and Nagel \cite{Oddershede:prl2000} used high speed photography to track the evolution of a sharp liquid spout formed in response to a large normal electric field applied to a layer of distilled water containing NaCl beneath a layer of silicone oil. Above a critical field strength, the water generated a cuspidal shape whose apex curvature and overall height were shown to exhibit divergent self-similar growth extending almost three decades in time. The  authors indicated that close to the collapse time, the shape of the conducting liquid tip was not exactly a cone but instead more  closely approximated a square-root cusp. Simulations of the type described in this work could be expanded to examine such systems by replacing the perfectly conducting liquid with a partially conducting one and replacing the vacuum layer with a viscous liquid layer. These changes, of course, will modify the requisite boundary conditions at the moving interface.

More recently, Elele \textit{et al.} \cite{Elele:prl2015} have been successful in monitoring the field driven deformation of submillimeter size sessile droplets into conical shapes under ambient conditions by limiting the buildup of electric charge along the free surface. This allowed a sufficiently long period of distortion and growth before the onset of electrospray, which was achieved by separating the liquid droplet from the ground electrode with an electrically insulating film that prevented electric current from flowing between opposing electrodes. They examined the behavior of a number of partially conducting and leaky dielectric liquids in both ground based and space based studies. Their microgravity experiments carried out in the International Space Station allowed measurements on even larger droplet sizes in the centimeter range. Elele \textit{et al.} confirmed the growth of conic tip shapes exhibiting a wide range of internal vertex angles dependent on operating conditions and self-similar behavior in the gap distance between the liquid apex and counter electrode, found to scale roughly as  $\tau^{-2/3}$. Although this behavior is suggestive of Zubarev's original prediction given by Eq. (\ref{eqn:StartSelfSim}), it is important to note that the liquids used in the experiments by Elele et al. were not perfectly conducting nor the flow regime controlled by inviscid conditions, both assumptions critical to Zubarev's findings.

\section{Conclusion}
The numerical simulations described in this work demonstrate how the  free surface of an electrified perfectly conducting viscous liquid like a liquid metal undergoes deformation to a rapidly accelerating  cusp with a conic tip for Reynolds numbers spanning the viscous to inertial regime. Field enhancement caused by the continuous sharpening of the conic apex causes divergent power law growth in finite time. Temporal tracking of the interface shape, normal stresses and forces per unit volume at the conic tip reveals persistent self-similar growth upon approach to the collapse time. The associated power laws exponents cover a dynamic range spanning six orders of magnitude in the Reynolds number. The asymptotic values of these exponents are in excellent agreement with previous predictions in the literature for the distinguished limits $Re = 0$ (Stokes regime) and $Re \rightarrow \infty$ (inviscid regime). The simulation results support the conclusion that at very small Reynolds number, the dominant balance at the tip as represented by the forces per unit volume acting on the conic apex represents the competition between the Maxwell and viscous forces while at intermediate and larger Reynolds number, the dominant balance switches to the competition between Maxwell and capillary stresses.

The temporal behavior of the terms in the normal stress boundary condition at the moving interface indicate that for Reynolds numbers  higher than about 50, the viscous normal stress is rather insignificant in comparison to the Maxwell and capillary pressure. Similar tracking of the terms in the Navier-Stokes equation, however, reveals that viscous normal forces per unit volume acting at the conic apex cannot be neglected for values \textsf{Re} below about $3 \times 10^4$ since they remain comparable in magnitude to the apical pressure gradient and apical inertial forces. Previous studies of dynamic cone formation based on Bernoulli's equation have tended to neglect altogether the role of viscous stresses at the cone tip.

Zubarev's \cite{Zubarev:jetp2001} original prediction for self-similar growth in the inviscid limit led to power law exponents characterized by a ratio of two integers. In particular, he reported that the capillary pressure, the Maxwell pressure and the kinetic energy per unit volume all scaled as $\tau^{-2/3}$. By contrast, Fontelos et al. \cite{Fontelos:pof2008} uncovered that in the Stokes regime the corresponding exponents are not simple integer ratios but instead depend functionally on the cone interior half-angle. Our study, which reveals the range and behavior of exponents characterizing the forces at the tip, confirm that the exponents at finite \textsf{Re} are never simple integer values. These non-integer ratios reflect the fact that the forces operating at the tip depend sensitively on the slope of the dynamic cone as the system approaches the collapse time.

In conclusion, the study presented here is based on examination of the forces operating at the conic tip and the resultant interface shapes which are shown to depend on the internal cone angle. We are currently expanding our study to evaluate the flow behavior throughout the tip region to better understand the characteristics of the interfacial boundary layer which accompanies the movement of the rapidly accelerating and highly curved interface. Such a boundary layer is expected to affect not only the stability of this interfacial region but ultimately the field emission process known to emanate from the conic apex.

\begin{acknowledgments}
The authors gratefully acknowledge financial support from a 2014 NASA Space Technology Research Fellowship (TGA) and the NASA/Jet Propulsion Laboratory (JPL) President's and Director's Fund (SMT). The authors also wish to acknowledge Chengzhe Zhou for interesting discussions during the course of this work about Taylor cone formation in the inviscid limit.
\end{acknowledgments}


\begin{thebibliography}{38}%
\makeatletter
\providecommand \@ifxundefined [1]{%
 \@ifx{#1\undefined}
}%
\providecommand \@ifnum [1]{%
 \ifnum #1\expandafter \@firstoftwo
 \else \expandafter \@secondoftwo
 \fi
}%
\providecommand \@ifx [1]{%
 \ifx #1\expandafter \@firstoftwo
 \else \expandafter \@secondoftwo
 \fi
}%
\providecommand \natexlab [1]{#1}%
\providecommand \enquote  [1]{``#1''}%
\providecommand \bibnamefont  [1]{#1}%
\providecommand \bibfnamefont [1]{#1}%
\providecommand \citenamefont [1]{#1}%
\providecommand \href@noop [0]{\@secondoftwo}%
\providecommand \href [0]{\begingroup \@sanitize@url \@href}%
\providecommand \@href[1]{\@@startlink{#1}\@@href}%
\providecommand \@@href[1]{\endgroup#1\@@endlink}%
\providecommand \@sanitize@url [0]{\catcode `\\12\catcode `\$12\catcode
  `\&12\catcode `\#12\catcode `\^12\catcode `\_12\catcode `\%12\relax}%
\providecommand \@@startlink[1]{}%
\providecommand \@@endlink[0]{}%
\providecommand \url  [0]{\begingroup\@sanitize@url \@url }%
\providecommand \@url [1]{\endgroup\@href {#1}{\urlprefix }}%
\providecommand \urlprefix  [0]{URL }%
\providecommand \Eprint [0]{\href }%
\providecommand \doibase [0]{http://dx.doi.org/}%
\providecommand \selectlanguage [0]{\@gobble}%
\providecommand \bibinfo  [0]{\@secondoftwo}%
\providecommand \bibfield  [0]{\@secondoftwo}%
\providecommand \translation [1]{[#1]}%
\providecommand \BibitemOpen [0]{}%
\providecommand \bibitemStop [0]{}%
\providecommand \bibitemNoStop [0]{.\EOS\space}%
\providecommand \EOS [0]{\spacefactor3000\relax}%
\providecommand \BibitemShut  [1]{\csname bibitem#1\endcsname}%
\let\auto@bib@innerbib\@empty
\bibitem [{\citenamefont {Larmor}(1890)}]{Larmor:pcps1890}%
  \BibitemOpen
  \bibfield  {author} {\bibinfo {author} {\bibfnamefont {J.}~\bibnamefont
  {Larmor}},\ }\bibfield  {title} {\enquote {\bibinfo {title} {On the influence
  of electrification on ripples},}\ }\href@noop {} {\bibfield  {journal}
  {\bibinfo  {journal} {Proc. Cambridge Phil. Soc.}\ }\textbf {\bibinfo
  {volume} {7}},\ \bibinfo {pages} {69--71} (\bibinfo {year}
  {1890})}\BibitemShut {NoStop}%
\bibitem [{\citenamefont {Tonks}(1935)}]{Tonks:pr1935}%
  \BibitemOpen
  \bibfield  {author} {\bibinfo {author} {\bibfnamefont {L.}~\bibnamefont
  {Tonks}},\ }\bibfield  {title} {\enquote {\bibinfo {title} {A theory of
  liquid surface rupture by a uniform electric field},}\ }\href@noop {}
  {\bibfield  {journal} {\bibinfo  {journal} {Phys. Rev}\ }\textbf {\bibinfo
  {volume} {48}},\ \bibinfo {pages} {562--568} (\bibinfo {year}
  {1935})}\BibitemShut {NoStop}%
\bibitem [{\citenamefont {Frenkel}(1935)}]{Frenkel:zetf1935}%
  \BibitemOpen
  \bibfield  {author} {\bibinfo {author} {\bibfnamefont {J.}~\bibnamefont
  {Frenkel}},\ }\bibfield  {title} {\enquote {\bibinfo {title} {On {`Tonks'}
  theory of liquid surface rupture by a uniform electric field},}\ }\href@noop
  {} {\bibfield  {journal} {\bibinfo  {journal} {Phys. Z. Sowjetunion}\
  }\textbf {\bibinfo {volume} {8}},\ \bibinfo {pages} {675 -- 683} (\bibinfo
  {year} {1935})}\BibitemShut {NoStop}%
\bibitem [{\citenamefont {Frenkel}(1936)}]{Frenkel:pzs1936}%
  \BibitemOpen
  \bibfield  {author} {\bibinfo {author} {\bibfnamefont {Ya.~I.}\ \bibnamefont
  {Frenkel}},\ }\bibfield  {title} {\enquote {\bibinfo {title} {About the
  {T}onks theory of liquid surface rupture by a uniform electric field in
  vacuum},}\ }\href@noop {} {\bibfield  {journal} {\bibinfo  {journal} {Zh.
  Eksp. Teor. Fiz}\ }\textbf {\bibinfo {volume} {6}},\ \bibinfo {pages}
  {347--50} (\bibinfo {year} {1936})}\BibitemShut {NoStop}%
\bibitem [{\citenamefont {Krohn}(1984)}]{Krohn:paa1961}%
  \BibitemOpen
  \bibfield  {author} {\bibinfo {author} {\bibfnamefont {V.~E.}\ \bibnamefont
  {Krohn}},\ }\bibfield  {title} {\enquote {\bibinfo {title} {Liquid metal
  droplets for heavy particle propulsion},}\ }\href@noop {} {\bibfield
  {journal} {\bibinfo  {journal} {Prog. Astronautics Aeronautics}\ }\textbf
  {\bibinfo {volume} {5}},\ \bibinfo {pages} {73--80} (\bibinfo {year}
  {1984})}\BibitemShut {NoStop}%
\bibitem [{\citenamefont {Perel}\ \emph {et~al.}(1971)\citenamefont {Perel},
  \citenamefont {Yahiku}, \citenamefont {Mahoney}, \citenamefont {Daley},\ and\
  \citenamefont {Sherman}}]{Perel:jsr1971}%
  \BibitemOpen
  \bibfield  {author} {\bibinfo {author} {\bibfnamefont {J.}~\bibnamefont
  {Perel}}, \bibinfo {author} {\bibfnamefont {A.~Y.}\ \bibnamefont {Yahiku}},
  \bibinfo {author} {\bibfnamefont {J.~F.}\ \bibnamefont {Mahoney}}, \bibinfo
  {author} {\bibfnamefont {H.~L.}\ \bibnamefont {Daley}}, \ and\ \bibinfo
  {author} {\bibfnamefont {A.}~\bibnamefont {Sherman}},\ }\bibfield  {title}
  {\enquote {\bibinfo {title} {Operational characteristics of colloid
  thrusters},}\ }\href@noop {} {\bibfield  {journal} {\bibinfo  {journal} {J.
  Spacecraft Rockets}\ }\textbf {\bibinfo {volume} {8}},\ \bibinfo {pages}
  {702--708} (\bibinfo {year} {1971})}\BibitemShut {NoStop}%
\bibitem [{\citenamefont {Bailey}\ \emph {et~al.}(1972)\citenamefont {Bailey},
  \citenamefont {Bracher},\ and\ \citenamefont {Von~Rohden}}]{Bailey:jsr1972}%
  \BibitemOpen
  \bibfield  {author} {\bibinfo {author} {\bibfnamefont {A.~G.}\ \bibnamefont
  {Bailey}}, \bibinfo {author} {\bibfnamefont {J.~E.}\ \bibnamefont {Bracher}},
  \ and\ \bibinfo {author} {\bibfnamefont {H.~J.}\ \bibnamefont {Von~Rohden}},\
  }\bibfield  {title} {\enquote {\bibinfo {title} {A capillary-fed annular
  colloid thruster},}\ }\href@noop {} {\bibfield  {journal} {\bibinfo
  {journal} {J. Spacecraft Rockets}\ }\textbf {\bibinfo {volume} {9}},\
  \bibinfo {pages} {518--521} (\bibinfo {year} {1972})}\BibitemShut {NoStop}%
\bibitem [{\citenamefont {Zafran}\ \emph {et~al.}(1973)\citenamefont {Zafran},
  \citenamefont {Beynon}, \citenamefont {Kidd}, \citenamefont {Shelton},\ and\
  \citenamefont {Jackson}}]{Zafran:jsr1973}%
  \BibitemOpen
  \bibfield  {author} {\bibinfo {author} {\bibfnamefont {S.}~\bibnamefont
  {Zafran}}, \bibinfo {author} {\bibfnamefont {J.C.}\ \bibnamefont {Beynon}},
  \bibinfo {author} {\bibfnamefont {P.W.}\ \bibnamefont {Kidd}}, \bibinfo
  {author} {\bibfnamefont {H.}~\bibnamefont {Shelton}}, \ and\ \bibinfo
  {author} {\bibfnamefont {F.~A}\ \bibnamefont {Jackson}},\ }\bibfield  {title}
  {\enquote {\bibinfo {title} {One-mlb colloid thruster system development},}\
  }\href@noop {} {\bibfield  {journal} {\bibinfo  {journal} {J. Spacecraft
  Rockets}\ }\textbf {\bibinfo {volume} {10}},\ \bibinfo {pages} {531--533}
  (\bibinfo {year} {1973})}\BibitemShut {NoStop}%
\bibitem [{\citenamefont {Huberman}\ and\ \citenamefont
  {Rosen}(1974)}]{Huberman:jsr1974}%
  \BibitemOpen
  \bibfield  {author} {\bibinfo {author} {\bibfnamefont {M.~N.}\ \bibnamefont
  {Huberman}}\ and\ \bibinfo {author} {\bibfnamefont {S.~G.}\ \bibnamefont
  {Rosen}},\ }\bibfield  {title} {\enquote {\bibinfo {title} {Advanced
  high-thrust colloid sources},}\ }\href@noop {} {\bibfield  {journal}
  {\bibinfo  {journal} {J. Spacecraft Rockets}\ }\textbf {\bibinfo {volume}
  {11}},\ \bibinfo {pages} {475 -- 480} (\bibinfo {year} {1974})}\BibitemShut
  {NoStop}%
\bibitem [{\citenamefont {Krohn}\ and\ \citenamefont
  {Ringo}(1972)}]{Krohn:rsi1972}%
  \BibitemOpen
  \bibfield  {author} {\bibinfo {author} {\bibfnamefont {V.~E.}\ \bibnamefont
  {Krohn}}\ and\ \bibinfo {author} {\bibfnamefont {G.~R.}\ \bibnamefont
  {Ringo}},\ }\bibfield  {title} {\enquote {\bibinfo {title} {Secondary-ion
  collection system for an ion microprobe analyzer of high mass resolution},}\
  }\href@noop {} {\bibfield  {journal} {\bibinfo  {journal} {Rev. Sci.
  Instrum.}\ }\textbf {\bibinfo {volume} {43}},\ \bibinfo {pages} {1771--1772}
  (\bibinfo {year} {1972})}\BibitemShut {NoStop}%
\bibitem [{\citenamefont {Seliger}\ \emph {et~al.}(1979)\citenamefont
  {Seliger}, \citenamefont {Ward}, \citenamefont {Wang},\ and\ \citenamefont
  {Kubena}}]{Seliger:apl1979}%
  \BibitemOpen
  \bibfield  {author} {\bibinfo {author} {\bibfnamefont {R.~L.}\ \bibnamefont
  {Seliger}}, \bibinfo {author} {\bibfnamefont {J.~W.}\ \bibnamefont {Ward}},
  \bibinfo {author} {\bibfnamefont {V.}~\bibnamefont {Wang}}, \ and\ \bibinfo
  {author} {\bibfnamefont {R.~L.}\ \bibnamefont {Kubena}},\ }\bibfield  {title}
  {\enquote {\bibinfo {title} {A high-intensity scanning ion probe with
  submicrometer spot size},}\ }\href@noop {} {\bibfield  {journal} {\bibinfo
  {journal} {Appl. Phys. Lett.}\ }\textbf {\bibinfo {volume} {34}},\ \bibinfo
  {pages} {310--312} (\bibinfo {year} {1979})}\BibitemShut {NoStop}%
\bibitem [{\citenamefont {Orloff}\ \emph {et~al.}(2003)\citenamefont {Orloff},
  \citenamefont {M.},\ and\ \citenamefont {Swanson}}]{Orloff:2003}%
  \BibitemOpen
  \bibfield  {author} {\bibinfo {author} {\bibfnamefont {J.}~\bibnamefont
  {Orloff}}, \bibinfo {author} {\bibfnamefont {Utlaut}\ \bibnamefont {M.}}, \
  and\ \bibinfo {author} {\bibfnamefont {L}~\bibnamefont {Swanson}},\
  }\href@noop {} {\emph {\bibinfo {title} {High Resolution Focused Ion Beams:
  FIB and Its Applications}}}\ (\bibinfo  {publisher} {Kluwer Academic},\
  \bibinfo {year} {2003})\BibitemShut {NoStop}%
\bibitem [{\citenamefont {Wright}\ and\ \citenamefont
  {Ferrer}(2015)}]{Wright:pas2015}%
  \BibitemOpen
  \bibfield  {author} {\bibinfo {author} {\bibfnamefont {W.~P.}\ \bibnamefont
  {Wright}}\ and\ \bibinfo {author} {\bibfnamefont {P.}~\bibnamefont
  {Ferrer}},\ }\bibfield  {title} {\enquote {\bibinfo {title} {Electric
  micropropulsionsystems},}\ }\href@noop {} {\bibfield  {journal} {\bibinfo
  {journal} {Prog. Aero. Sci.}\ }\textbf {\bibinfo {volume} {74}},\ \bibinfo
  {pages} {48--61} (\bibinfo {year} {2015})}\BibitemShut {NoStop}%
\bibitem [{\citenamefont {Bharti}\ and\ \citenamefont
  {Chalia}(2017)}]{Bharti:irjet2017}%
  \BibitemOpen
  \bibfield  {author} {\bibinfo {author} {\bibfnamefont {M.~K.}\ \bibnamefont
  {Bharti}}\ and\ \bibinfo {author} {\bibfnamefont {S.}~\bibnamefont
  {Chalia}},\ }\bibfield  {title} {\enquote {\bibinfo {title} {Literature study
  of field emission electric propulsion microthruster},}\ }\href@noop {}
  {\bibfield  {journal} {\bibinfo  {journal} {Inter. Res. J. Eng. and Tech.}\
  }\textbf {\bibinfo {volume} {4}},\ \bibinfo {pages} {2777--2781} (\bibinfo
  {year} {2017})}\BibitemShut {NoStop}%
\bibitem [{NAS(2015)}]{NASA15}%
  \BibitemOpen
  \href@noop {} {\enquote {\bibinfo {title} {{NASA Technology Roadmaps TA2}:
  {In-Space Propulsion Technologies}},}\ }\bibinfo {howpublished}
  {www.nasa.gov/offices/oct/home/roadmaps/index.html} (\bibinfo {year}
  {2015})\BibitemShut {NoStop}%
\bibitem [{\citenamefont {You}(2017)}]{You:2018}%
  \BibitemOpen
  \bibfield  {author} {\bibinfo {author} {\bibfnamefont {Zheng}\ \bibnamefont
  {You}},\ }\href@noop {} {\emph {\bibinfo {title} {Space Microsystems and
  Micro/Nano Satellites}}},\ National Defense Industry Press\ (\bibinfo
  {publisher} {Butterworth-Heinemann},\ \bibinfo {address} {Oxford, UK},\
  \bibinfo {year} {2017})\BibitemShut {NoStop}%
\bibitem [{\citenamefont {Rayleigh}(1882)}]{Rayleigh:pm1882}%
  \BibitemOpen
  \bibfield  {author} {\bibinfo {author} {\bibfnamefont {{Lord}}\ \bibnamefont
  {Rayleigh}},\ }\bibfield  {title} {\enquote {\bibinfo {title} {On the
  equilibrium of liquid conducting masses charged with electricity},}\
  }\href@noop {} {\bibfield  {journal} {\bibinfo  {journal} {Philos. Mag.}\
  }\textbf {\bibinfo {volume} {14}},\ \bibinfo {pages} {184--186} (\bibinfo
  {year} {1882})}\BibitemShut {NoStop}%
\bibitem [{\citenamefont {Taylor}(1964)}]{Taylor:prsla1964}%
  \BibitemOpen
  \bibfield  {author} {\bibinfo {author} {\bibfnamefont {G.~I.}\ \bibnamefont
  {Taylor}},\ }\bibfield  {title} {\enquote {\bibinfo {title} {Disintegration
  of water drops in an electric field},}\ }\href@noop {} {\bibfield  {journal}
  {\bibinfo  {journal} {Proc. R. Soc. Lond. A}\ }\textbf {\bibinfo {volume}
  {280}},\ \bibinfo {pages} {383--397} (\bibinfo {year} {1964})}\BibitemShut
  {NoStop}%
\bibitem [{\citenamefont {Gilbert}()}]{Gilbert:petrus1600}%
  \BibitemOpen
  \bibfield  {author} {\bibinfo {author} {\bibfnamefont {W.}~\bibnamefont
  {Gilbert}},\ }\href@noop {} {\enquote {\bibinfo {title} {De {M}agnete},}\
  }\bibinfo {howpublished} {(Petrus Short, London, 1600 (in Latin)).
  Translation by P. F. Mottlay, pg. 89 (Dover, New York, 1958)}\BibitemShut
  {NoStop}%
\bibitem [{\citenamefont {Gray}(1731)}]{Gray:prsl1731}%
  \BibitemOpen
  \bibfield  {author} {\bibinfo {author} {\bibfnamefont {S.}~\bibnamefont
  {Gray}},\ }\bibfield  {title} {\enquote {\bibinfo {title} {{II.} {A} letter
  concerning the electricity of water},}\ }\href@noop {} {\bibfield  {journal}
  {\bibinfo  {journal} {Proc. R. Soc. Lond.}\ }\textbf {\bibinfo {volume}
  {37}},\ \bibinfo {pages} {227--230 {and} 260} (\bibinfo {year}
  {1731})}\BibitemShut {NoStop}%
\bibitem [{\citenamefont {Zubarev}(2001)}]{Zubarev:jetp2001}%
  \BibitemOpen
  \bibfield  {author} {\bibinfo {author} {\bibfnamefont {N.~M.}\ \bibnamefont
  {Zubarev}},\ }\bibfield  {title} {\enquote {\bibinfo {title} {Formation of
  conic cusps at the surface of liquid metal in electric field},}\ }\href@noop
  {} {\bibfield  {journal} {\bibinfo  {journal} {J. Exp. Theor. Phys.}\
  }\textbf {\bibinfo {volume} {30}},\ \bibinfo {pages} {544--548} (\bibinfo
  {year} {2001})}\BibitemShut {NoStop}%
\bibitem [{\citenamefont {Suvorov}\ and\ \citenamefont
  {Litvinov}(2000)}]{Suvorov:jpd2000}%
  \BibitemOpen
  \bibfield  {author} {\bibinfo {author} {\bibfnamefont {V.~G.}\ \bibnamefont
  {Suvorov}}\ and\ \bibinfo {author} {\bibfnamefont {E.~A.}\ \bibnamefont
  {Litvinov}},\ }\bibfield  {title} {\enquote {\bibinfo {title} {Dynamic taylor
  cone formation on liquid metal surface: numerical modeling},}\ }\href@noop {}
  {\bibfield  {journal} {\bibinfo  {journal} {J. Phys. D: Appl. Phys.}\
  }\textbf {\bibinfo {volume} {33}},\ \bibinfo {pages} {1245--1251} (\bibinfo
  {year} {2000})}\BibitemShut {NoStop}%
\bibitem [{\citenamefont {Belozerkovskii}(1994)}]{Belozerkovskii:1994}%
  \BibitemOpen
  \bibfield  {author} {\bibinfo {author} {\bibfnamefont {O.~M.}\ \bibnamefont
  {Belozerkovskii}},\ }\href@noop {} {\emph {\bibinfo {title} {Numerical
  Simulations in Continuous Media Mechanics}}}\ (\bibinfo  {publisher} {Physics
  and Mathematics Literature},\ \bibinfo {address} {Moscow},\ \bibinfo {year}
  {1994})\ p.\ \bibinfo {pages} {442}\BibitemShut {NoStop}%
\bibitem [{\citenamefont {Zheng}\ and\ \citenamefont
  {Linsu}(1988)}]{Cui:jvstb1988}%
  \BibitemOpen
  \bibfield  {author} {\bibinfo {author} {\bibfnamefont {C.}~\bibnamefont
  {Zheng}}\ and\ \bibinfo {author} {\bibfnamefont {T.}~\bibnamefont {Linsu}},\
  }\bibfield  {title} {\enquote {\bibinfo {title} {A new approach to simulating
  the operation of liquid metal ion sources},}\ }\href@noop {} {\bibfield
  {journal} {\bibinfo  {journal} {J. Vac. Sci. \& Tech.}\ }\textbf {\bibinfo
  {volume} {6}},\ \bibinfo {pages} {2104 -- 2107} (\bibinfo {year}
  {1988})}\BibitemShut {NoStop}%
\bibitem [{\citenamefont {Suvorov}(2004)}]{Suvorov:surf2004}%
  \BibitemOpen
  \bibfield  {author} {\bibinfo {author} {\bibfnamefont {V.~G.}\ \bibnamefont
  {Suvorov}},\ }\bibfield  {title} {\enquote {\bibinfo {title} {Numerical
  analysis of liquid metal flow in the presence of an electric
  field:application to liquid metal ion source},}\ }\href@noop {} {\bibfield
  {journal} {\bibinfo  {journal} {Surf. Interface Anal.}\ }\textbf {\bibinfo
  {volume} {36}},\ \bibinfo {pages} {421--425} (\bibinfo {year}
  {2004})}\BibitemShut {NoStop}%
\bibitem [{\citenamefont {Suvorov}\ and\ \citenamefont
  {Zubarev}(2004)}]{Suvorov:jpd2004}%
  \BibitemOpen
  \bibfield  {author} {\bibinfo {author} {\bibfnamefont {V.~G.}\ \bibnamefont
  {Suvorov}}\ and\ \bibinfo {author} {\bibfnamefont {N.~M.}\ \bibnamefont
  {Zubarev}},\ }\bibfield  {title} {\enquote {\bibinfo {title} {Formation of
  the taylor cone on the surface of liquid metal in the presence of an electric
  field},}\ }\href@noop {} {\bibfield  {journal} {\bibinfo  {journal} {J. Phys.
  D: Appl. Phys.}\ }\textbf {\bibinfo {volume} {37}},\ \bibinfo {pages}
  {289--297} (\bibinfo {year} {2004})}\BibitemShut {NoStop}%
\bibitem [{\citenamefont {Collins}\ \emph {et~al.}(2008)\citenamefont
  {Collins}, \citenamefont {Jones}, \citenamefont {Harris},\ and\ \citenamefont
  {Basaran}}]{Collins:natp2008}%
  \BibitemOpen
  \bibfield  {author} {\bibinfo {author} {\bibfnamefont {R.~T.}\ \bibnamefont
  {Collins}}, \bibinfo {author} {\bibfnamefont {J.~J.}\ \bibnamefont {Jones}},
  \bibinfo {author} {\bibfnamefont {M.~T.}\ \bibnamefont {Harris}}, \ and\
  \bibinfo {author} {\bibfnamefont {O.~A.}\ \bibnamefont {Basaran}},\
  }\bibfield  {title} {\enquote {\bibinfo {title} {Electrohydrodynamic tip
  streaming and emission of charged drops from liquid cones},}\ }\href@noop {}
  {\bibfield  {journal} {\bibinfo  {journal} {Nature Physics}\ }\textbf
  {\bibinfo {volume} {4}},\ \bibinfo {pages} {149--154} (\bibinfo {year}
  {2008})}\BibitemShut {NoStop}%
\bibitem [{\citenamefont {de~la Mora}\ and\ \citenamefont
  {Loscertales}(1994)}]{delaMora:jfm1994}%
  \BibitemOpen
  \bibfield  {author} {\bibinfo {author} {\bibfnamefont {J.~F.}\ \bibnamefont
  {de~la Mora}}\ and\ \bibinfo {author} {\bibfnamefont {I.}~\bibnamefont
  {Loscertales}},\ }\bibfield  {title} {\enquote {\bibinfo {title} {The current
  emitted by highly conducting taylor cones},}\ }\href@noop {} {\bibfield
  {journal} {\bibinfo  {journal} {J. Fluid Mech.}\ }\textbf {\bibinfo {volume}
  {260}},\ \bibinfo {pages} {155--184} (\bibinfo {year} {1994})}\BibitemShut
  {NoStop}%
\bibitem [{\citenamefont {Ga\~n\'an{-}Calvo}(2001)}]{GananCalvo:prl1997}%
  \BibitemOpen
  \bibfield  {author} {\bibinfo {author} {\bibfnamefont {A.~M.}\ \bibnamefont
  {Ga\~n\'an{-}Calvo}},\ }\bibfield  {title} {\enquote {\bibinfo {title}
  {Electrospray as a source of nanoparticles for efficient colloid
  thrusters},}\ }\href@noop {} {\bibfield  {journal} {\bibinfo  {journal}
  {Phys. Rev. Lett.}\ }\textbf {\bibinfo {volume} {17}},\ \bibinfo {pages}
  {217--220} (\bibinfo {year} {2001})}\BibitemShut {NoStop}%
\bibitem [{\citenamefont {Burton}\ and\ \citenamefont
  {Taborek}(2011)}]{Burton:prl2011}%
  \BibitemOpen
  \bibfield  {author} {\bibinfo {author} {\bibfnamefont {J.~C.}\ \bibnamefont
  {Burton}}\ and\ \bibinfo {author} {\bibfnamefont {P.}~\bibnamefont
  {Taborek}},\ }\bibfield  {title} {\enquote {\bibinfo {title} {Simulations of
  coulombic fission of charged inviscid drops},}\ }\href@noop {} {\bibfield
  {journal} {\bibinfo  {journal} {Phys. Rev. Lett.}\ }\textbf {\bibinfo
  {volume} {106}},\ \bibinfo {pages} {144501} (\bibinfo {year}
  {2011})}\BibitemShut {NoStop}%
\bibitem [{\citenamefont {Garzon}\ \emph {et~al.}(2014)\citenamefont {Garzon},
  \citenamefont {Gray},\ and\ \citenamefont {Sethian}}]{Garzon:pre2014}%
  \BibitemOpen
  \bibfield  {author} {\bibinfo {author} {\bibfnamefont {M.}~\bibnamefont
  {Garzon}}, \bibinfo {author} {\bibfnamefont {L.~J.}\ \bibnamefont {Gray}}, \
  and\ \bibinfo {author} {\bibfnamefont {J.~A.}\ \bibnamefont {Sethian}},\
  }\bibfield  {title} {\enquote {\bibinfo {title} {Numerical simulations of
  electrostatically driven jets from nonviscous droplets},}\ }\href@noop {}
  {\bibfield  {journal} {\bibinfo  {journal} {Phys. Rev. E}\ }\textbf {\bibinfo
  {volume} {89}},\ \bibinfo {pages} {033011} (\bibinfo {year}
  {2014})}\BibitemShut {NoStop}%
\bibitem [{\citenamefont {Betel\'u}\ \emph {et~al.}(2006)\citenamefont
  {Betel\'u}, \citenamefont {Fontelos}, \citenamefont {Kindel\'an},\ and\
  \citenamefont {Vantzos}}]{Betelu:pof2006}%
  \BibitemOpen
  \bibfield  {author} {\bibinfo {author} {\bibfnamefont {S.~I.}\ \bibnamefont
  {Betel\'u}}, \bibinfo {author} {\bibfnamefont {M.~A.}\ \bibnamefont
  {Fontelos}}, \bibinfo {author} {\bibfnamefont {U.}~\bibnamefont
  {Kindel\'an}}, \ and\ \bibinfo {author} {\bibfnamefont {O.}~\bibnamefont
  {Vantzos}},\ }\bibfield  {title} {\enquote {\bibinfo {title} {Singularities
  on charged viscous droplets},}\ }\href@noop {} {\bibfield  {journal}
  {\bibinfo  {journal} {Phys. Fluids}\ }\textbf {\bibinfo {volume} {18}},\
  \bibinfo {pages} {051706} (\bibinfo {year} {2006})}\BibitemShut {NoStop}%
\bibitem [{\citenamefont {Fontelos}\ \emph {et~al.}(2008)\citenamefont
  {Fontelos}, \citenamefont {Kindel\'an},\ and\ \citenamefont
  {Vantzos}}]{Fontelos:pof2008}%
  \BibitemOpen
  \bibfield  {author} {\bibinfo {author} {\bibfnamefont {M.~A.}\ \bibnamefont
  {Fontelos}}, \bibinfo {author} {\bibfnamefont {U.}~\bibnamefont
  {Kindel\'an}}, \ and\ \bibinfo {author} {\bibfnamefont {O.}~\bibnamefont
  {Vantzos}},\ }\bibfield  {title} {\enquote {\bibinfo {title} {Evolution of
  neutral and charged droplets in an electric field},}\ }\href@noop {}
  {\bibfield  {journal} {\bibinfo  {journal} {Phys. Fluids}\ ,\ \bibinfo
  {pages} {092110}} (\bibinfo {year} {2008})}\BibitemShut {NoStop}%
\bibitem [{\citenamefont {COMSOL~Multiphysics}()}]{COMSOL}%
  \BibitemOpen
  \bibfield  {author} {\bibinfo {author} {\bibfnamefont {Inc.}\ \bibnamefont
  {COMSOL~Multiphysics}},\ }\href@noop {} {\enquote {\bibinfo {title}
  {Microfluidics module v5.2a},}\ }\bibinfo {howpublished} {Burlington, MA,
  USA}\BibitemShut {NoStop}%
\bibitem [{\citenamefont {Forbes}\ and\ \citenamefont
  {Mair}(2008)}]{Orloff:chapter2008}%
  \BibitemOpen
  \bibfield  {author} {\bibinfo {author} {\bibfnamefont {R.~G.}\ \bibnamefont
  {Forbes}}\ and\ \bibinfo {author} {\bibfnamefont {G.~L.~R.}\ \bibnamefont
  {Mair}},\ }\bibfield  {title} {\enquote {\bibinfo {title} {Liquid metal ion
  sources},}\ }in\ \href@noop {} {\emph {\bibinfo {booktitle} {Handbook of
  charged particle optics}}},\ \bibinfo {editor} {edited by\ \bibinfo {editor}
  {\bibfnamefont {J.}~\bibnamefont {Orloff}}}\ (\bibinfo  {publisher} {CRC
  Press},\ \bibinfo {address} {Boca Raton, FL},\ \bibinfo {year} {2008})\
  \bibinfo {edition} {2nd}\ ed.,\ Chap.~\bibinfo {chapter} {2}\BibitemShut
  {NoStop}%
\bibitem [{\citenamefont {Oddershede}\ and\ \citenamefont
  {Nagel}(2000)}]{Oddershede:prl2000}%
  \BibitemOpen
  \bibfield  {author} {\bibinfo {author} {\bibfnamefont {L.}~\bibnamefont
  {Oddershede}}\ and\ \bibinfo {author} {\bibfnamefont {S.~R.}\ \bibnamefont
  {Nagel}},\ }\bibfield  {title} {\enquote {\bibinfo {title} {Singularity
  during the onset of an electrohydrodynamic spout},}\ }\href@noop {}
  {\bibfield  {journal} {\bibinfo  {journal} {Phys. Rev. Lett.}\ }\textbf
  {\bibinfo {volume} {85}},\ \bibinfo {pages} {1234--1237} (\bibinfo {year}
  {2000})}\BibitemShut {NoStop}%
\bibitem [{\citenamefont {Elele}\ \emph {et~al.}(2015)\citenamefont {Elele},
  \citenamefont {Shen}, \citenamefont {Pettit},\ and\ \citenamefont
  {Khusid}}]{Elele:prl2015}%
  \BibitemOpen
  \bibfield  {author} {\bibinfo {author} {\bibfnamefont {E.~O.}\ \bibnamefont
  {Elele}}, \bibinfo {author} {\bibfnamefont {Y.}~\bibnamefont {Shen}},
  \bibinfo {author} {\bibfnamefont {D.~R.}\ \bibnamefont {Pettit}}, \ and\
  \bibinfo {author} {\bibfnamefont {B.}~\bibnamefont {Khusid}},\ }\bibfield
  {title} {\enquote {\bibinfo {title} {Detection of a dynamic cone-shaped
  meniscus on the surface of fluids in electric fields},}\ }\href@noop {}
  {\bibfield  {journal} {\bibinfo  {journal} {Phys. Rev. Lett.}\ }\textbf
  {\bibinfo {volume} {114}},\ \bibinfo {pages} {054501} (\bibinfo {year}
  {2015})}\BibitemShut {NoStop}%
\bibitem [{\citenamefont {Iida}\ and\ \citenamefont
  {Guthrie}(2007)}]{Iida:1988}%
  \BibitemOpen
  \bibfield  {author} {\bibinfo {author} {\bibfnamefont {T.}~\bibnamefont
  {Iida}}\ and\ \bibinfo {author} {\bibfnamefont {R.~I.~L.}\ \bibnamefont
  {Guthrie}},\ }\href@noop {} {\emph {\bibinfo {title} {The Physical Properties
  of Liquid Metals}}}\ (\bibinfo  {publisher} {Oxford University Press},\
  \bibinfo {year} {2007})\BibitemShut {NoStop}%
\end{thebibliography}
%

\begin{turnpage}
\begin{table*}[ht]
\renewcommand{\arraystretch}{0.7}
\caption{\label{table:betas} Data extracted from numerical simulations described in the text for $0.1 \leq \textsf{Re} \leq 50,000$ at $\textsf{Ca} = 7.0834$. Description of mesh schemes A and B appears in Section \ref{NumericalSimulation}. Definition of $\textsf{Re}$ is provided in Table \ref{table:nd}. Dimensionless times $T_f$ and $T_\textsf{C}$ denote simulation termination time and collapse time, as defined in the text. Variables $Z_f$ and $Z_C$ denote the corresponding vertical coordinates of the conic apex point. Variable $V_{apex,f}$ denotes the vertical flow speed at the conic apex at $T_f$. Power law exponents for the normal stresses at the conic apex diverge as $\tau^{-\beta_j}$ where $j=M, C \,\textrm{and}\,V$, referring to the Maxwell stress, capillary stress and viscous normal stress. Exponents $\beta_k$ for $k=1 - 4$ represent divergent growth of the four terms in Eq. (\ref {eqn:NDRzeroNS}) evaluated at the conic apex, which scale as $\tau^{-\beta_k}$. For all exponents listed in this table, the number appearing in parentheses represents the standard deviation multiplied by $10^{-4}$ (e.g. $0.7849(17)=0.7849\pm0.0017$).}
\newcolumntype{d}[1]{D{.}{.}{#1} }
\newcolumntype{t}[1]{D{x}{\times}{#1} }
\begin{tabular}{t{0} c d{6} d{6} d{6} d{6} d{4} d{8} d{8} d{8} d{8} d{8} d{8} d{8}}
\hline\hline
\noalign{\vskip 1.5mm}
\multicolumn{1}{l}{$~~~~\textsf{Re}$}  &
\multicolumn{1}{c}{Mesh} &
\multicolumn{1}{l}{$~~~T_f$} &  \multicolumn{1}{l}{$~~~T_\textsf{C}$}  &  \multicolumn{1}{l}{$~~~Z_f$}  &  \multicolumn{1}{l}{$~~~Z_\textsf{C}$}  &  \multicolumn{1}{l}{~$V_{apex,f}$}  &  \multicolumn{1}{l}{~~$\beta_M$}  &  \multicolumn{1}{l}{~~$\beta_C$}  &  \multicolumn{1}{l}{~~$\beta_V$} & \multicolumn{1}{l}{~~$\beta_1$}    &     \multicolumn{1}{l}{~~$\beta_2$} & \multicolumn{1}{l}{~~$\beta_3$} & \multicolumn{1}{l}{~~$\beta_4$}  \\
\noalign{\vskip 1.5mm}
\hline
\noalign{\vskip 1.5mm}
0.1      &A      &33.86686  &33.86825  &1.26818  &1.26961  &0.55       &0.9897(2)  &0.7849(17)      &1.0064(9)      &1.2735(21)       &1.3739(33)      &1.7815(39)      &1.7849(28)  \\
0.2      &A      &16.94872  &16.94941  &1.26917  &1.27068  &1.16       &0.9916(1)  &0.7773(13)      &1.0083(4)      &1.2856(17)       &1.3836(28)      &1.7798(24)      &1.7817(22)  \\
0.5      &A      &6.80949      &6.80975  &1.26970      &1.27113  &3.13       &0.9924(1)  &0.7616(2)      &1.0105(2)      &1.2960(13)       &1.3891(23)      &1.7744(8)      &1.7758(10)  \\
1      &A      &3.45222      &3.45235  &1.26663      &1.26807  &6.17       &0.9900(1)  &0.7593(3)      &1.0086(2)      &1.2981(14)       &1.3905(23)      &1.7679(16)      &1.7696(13)  \\
2      &A      &1.80664      &1.80671  &1.25564      &1.25708  &11.28       &0.9875(0)  &0.7562(2)      &1.0102(2)      &1.3080(14)       &1.3988(22)      &1.7636(10)      &1.7651(10)  \\
5      &A      &0.86960      &0.86964  &1.22165      &1.22296  &19.85       &0.9756(0)  &0.7558(2)      &1.0144(5)      &1.3239(15)       &1.4146(24)      &1.7517(20)      &1.7562(14)  \\
5      &B      &0.86922      &0.86926  &1.22160      &1.22290  &20.14       &0.9752(0)  &0.7517(1)      &1.0107(6)      &1.3219(16)       &1.4093(26)      &1.7413(3)      &1.7333(7)  \\
10      &A      &0.57190      &0.57192  &1.18893      &1.19010  &25.66       &0.9580(1)  &0.7529(3)      &1.0251(9)      &1.3412(16)       &1.4351(26)      &1.7388(15)      &1.7470(12)  \\
10      &B      &0.57176      &0.57178  &1.18896      &1.19009  &26.16       &0.9575(1)  &0.7498(3)      &1.0213(10)      &1.3389(17)       &1.4302(28)      &1.7275(6)      &1.7232(10)  \\
25      &A      &0.39252      &0.39253  &1.15385      &1.15474  &30.68       &0.9156(3)  &0.7463(6)      &1.0574(20)      &1.3600(19)       &1.4751(35)      &1.7035(12)      &1.7195(16)  \\
25      &B      &0.39230      &0.39232  &1.15379      &1.15470  &30.37       &0.9154(3)  &0.7459(6)      &1.0564(21)      &1.3594(20)       &1.4749(37)      &1.6976(11)      &1.7053(18)  \\
50      &A      &0.32952      &0.32953  &1.13632      &1.13701  &32.05       &0.8714(4)  &0.7392(7)      &1.1154(35)      &1.3697(20)       &1.5307(48)      &1.6677(13)      &1.6996(19)  \\
50      &B      &0.32928      &0.32930  &1.13625      &1.13694  &31.88       &0.8712(4)  &0.7392(8)      &1.1169(37)      &1.3701(21)       &1.5336(51)      &1.6628(13)      &1.6863(21)  \\
100  &A      &0.29560      &0.29560  &1.12514      &1.12567  &31.80       &0.8212(4)  &0.7288(8)      &\text{--}      &1.3758(21)       &\text{--}      &1.6209(14)      &1.6815(23)  \\
100  &B      &0.29545      &0.29546  &1.12508      &1.12561  &31.75       &0.8207(4)  &0.7274(9)      &\text{--}      &1.3733(22)       &\text{--}      &1.6139(15)      &1.6637(26)  \\
184.88  &B  &0.27859  &0.27859  &1.11890  &1.11933  &31.33       &0.7806(4)  &0.7186(8)      &\text{--}      &1.3804(19)       &\text{--}      &1.5679(13)      &1.6542(31)  \\
200  &B      &0.27700      &0.27701  &1.11830      &1.11871  &31.14       &0.7759(4)  &0.7167(8)      &\text{--}      &1.3792(20)       &\text{--}      &1.5607(12)      &1.6496(29)  \\
500  &B      &0.26482      &0.26483  &1.11338      &1.11371  &30.69       &0.7299(2)  &0.7054(6)      &\text{--}      &1.3891(11)       &\text{--}      &1.4804(15)      &1.5487(52)  \\
1000  &B  &0.26034      &0.26034  &1.11141      &1.11171  &30.08       &0.7092(1)  &0.7005(4)      &\text{--}      &1.3932(5)       &\text{--}      &1.4214(10)      &1.5594(43)  \\
2000  &B  &0.25791      &0.25792  &1.11026      &1.11055  &29.46       &0.6975(0)  &0.6975(3)      &0.7647(26)      &1.3922(3)       &1.1437(33)      &1.3821(9)      &1.5715(39)  \\
5000  &B  &0.25635      &0.25635  &1.10947      &1.10974  &29.18       &0.6892(1)  &0.6955(3)      &0.8759(10)      &1.3816(9)       &1.2530(16)      &1.3538(20)      &1.5671(51)  \\
10000  &B  &0.25578  &0.25578  &1.10916      &1.10943  &29.07       &0.6855(1)  &0.6935(3)      &0.9168(7)      &1.3694(11)       &1.2902(12)      &1.3399(8)      &1.5574(70)  \\
20000  &B  &0.25548  &0.25548  &1.10899      &1.10925  &28.83       &0.6835(1)  &0.6929(3)      &0.9448(18)      &1.3536(19)       &1.3170(20)      &1.3347(11)      &1.5582(80)  \\
25000  &B  &0.25537  &0.25538  &1.10895      &1.10921  &28.78       &0.6829(1)  &0.6926(3)      &0.9498(14)      &1.3491(21)       &1.3215(16)      &1.3344(11)      &1.5125(50)  \\
30000  &B  &0.25532  &0.25532  &1.10892      &1.10918  &28.72       &0.6825(1)  &0.6922(3)      &0.9568(12)      &1.3428(18)       &1.3274(14)      &1.3322(6)      &1.5111(61)  \\
40000  &B  &0.25524  &0.25524  &1.10888      &1.10915  &28.70       &0.6822(1)  &0.6925(2)      &0.9663(16)      &1.3296(22)       &1.3383(17)      &1.3302(7)      &1.4967(65)  \\
50000  &B  &0.25516  &0.25516  &1.10887      &1.10912  &29.36       &0.6817(1)  &0.6921(3)      &0.9719(25)      &1.3228(28)       &1.3430(25)      &1.3294(8)      &1.4893(69)  \\
\noalign{\vskip 1.5mm}
\hline
\end{tabular}
\end{table*}
\end{turnpage}

\begin{table*}[ht]
\caption{\label{table:materials} Relevant material constants and operating values for three liquid metals - Ga, In, and Cs - typically used in liquid metal ion sources. Ga values from Ref. [\onlinecite{Suvorov:surf2004}]. In and Cs values from Ref. [\onlinecite{Iida:1988}]. Quantities $\gamma$, $\mu$ and $\rho$ denote respectively the values of liquid surface tension, viscosity and density evaluated at the melting point. Definitions of $\textsf{Re}$, $\textsf{Ca}$ and $\textsf{Bo}$ can be found in Table \ref{table:nd}. Remaining quantities refer to the characteristic quantities defined in Table \ref{table:nd}, namely the voltage potential $\phi_o$, the length scale $h_o$ and the electric field strength $E_o$.}
\newcolumntype{t}[1]{D{x}{\times}{#1} }
\begin{tabular}{c c t{0.1} t{0.0} | c c | t{0.3} t{0.3} t{0.3} t{0.3}}
\hline
\hline
\noalign{\vskip 1.5mm}
\multicolumn{1}{m{0.1cm}}{ }  &  \multicolumn{1}{l}{~~$\gamma$\text{ (N/m)}~~~~} &  \multicolumn{1}{l}{~$\mu$\text{ (Pa}$\cdot$\text{s)}~~~} &  \multicolumn{1}{l}{~$\rho$\text{ (kg}$/$\text{m}$^3$)~~~}  &  \multicolumn{1}{l}{~~~~$\textsf{Re}$~~~~}  &  \multicolumn{1}{l}{$~~~~\textsf{Ca}~~~~$}  &  \multicolumn{1}{l}{$~~~~\phi_o~(V)~~~~~~~$}  &  \multicolumn{1}{l}{$h_o~(m)~~$}  &  \multicolumn{1}{l}{$~~~E_o~(V/m)~~~~~~$}  &  \multicolumn{1}{l}{$~Bo~~~~~$} \\
\noalign{\vskip 1.5mm}
\hline
\noalign{\vskip 1.5mm}
Ga &0.720    &2.132x10^{-3}    &6.09x10^{3}    &0.1        &7.0834     &1.30x10^{0}    &1.46x10^{-12}        &8.87x10^{11}    &1.77x10^{-19}     \\
       &     &            &            &50        &7.0834    &6.49x10^{2}     &3.66x10^{-7}        &1.77x10^{9}    &1.11x10^{-8}\\
       &     &            &            &500        &7.0834    &6.49x10^{3}     &3.66x10^{-5}        &1.77x10^{8}    &1.11x10^{-4}\\
       &     &            &            &5000        &7.0834    &6.49x10^{4}     &3.66x10^{-3}        &1.77x10^{7}    &1.11x10^{0}\\
       &     &            &            &50\,000    &7.0834    &6.49x10^{5}     &3.66x10^{-1}        &1.77x10^{6}    &1.11x10^{4}\\

\noalign{\vskip 0.8mm}
\hline
\noalign{\vskip 1.5mm}
In &0.556    &1.80x10^{-3}    &7.03x10^{3}    &0.1        &7.0834     &1.02x10^{0}    &1.17x10^{-12}        &8.72x10^{11}    &1.70x10^{-19}     \\
       &     &             &             &50        &7.0834    &5.10x10^{2}     &2.93x10^{-7}        &1.74x10^{9}    &1.06x10^{-8}\\
       &     &             &             &500        &7.0834     &5.10x10^{3}    &2.93x10^{-5}        &1.74x10^{8}     &1.06x10^{-4}\\
       &     &             &             &5000        &7.0834     &5.10x10^{4}    &2.93x10^{-3}        &1.74x10^{7}     &1.06x10^{0}\\
       &     &             &             &50\,000    &7.0834    &5.10x10^{5}     &2.93x10^{-1}        &1.74x10^{6}    &1.06x10^{4}\\
\noalign{\vskip 0.8mm}
\hline
\noalign{\vskip 1.5mm}
Cs &0.070    &6.86x10^{-4}    &1.84x10^{3}    &0.1        &7.0834     &7.60x10^{-1}    &5.16x10^{-12}        &1.47x10^{11}    &6.86x10^{-18}     \\
       &        &            &            &50        &7.0834    &3.80x10^{2}     &1.29x10^{-6}        &2.95x10^{8}    &4.29x10^{-7}\\
       &        &            &            &500        &7.0834     &3.80x10^{3}    &1.29x10^{-4}        &2.95x10^{7}     &4.29x10^{-3}\\
       &        &            &            &5000        &7.0834     &3.80x10^{4}    &1.29x10^{-2}        &2.95x10^{6}     &4.29x10^{1}\\
       &        &            &            &50\,000    &7.0834     &3.80x10^{5}    &1.29x10^{0}        &2.95x10^{5}     &4.29x10^{5}\\
\noalign{\vskip 0.8mm}
\hline
\end{tabular}
\end{table*}

\end{document}